\documentclass[useAMS,usenatbib]{mn2e}
\usepackage{myaasmacros}
\usepackage{graphicx}
\usepackage{ulem}
\usepackage{amsmath}
\usepackage{amssymb}

\def\ltsima{$\; \buildrel < \over \sim \;$}
\def\simlt{\lower.5ex\hbox{\ltsima}}   
\def\gtsima{$\; \buildrel > \over \sim \;$}
\def\simgt{\lower.5ex\hbox{\gtsima}}

\newcommand{\pc} {{\,\rm pc}} 
 
\newcommand{\cc}{{\,\rm {cm^{-3}}}}

\def\Myr{\,{\rm Myr}}
 
\newcommand{\kmsec}{{\,\rm {km\,s^{-1}} }}
\def\HH{{{\rm H}_2}}
\def\Msun{\, M_{\odot}}

\def\coreNFW{{\sc coreNFW}}

\def\Mtwodm{M_{200,{\rm DM}}}

\title[Dark matter cores all the way down]{Dark matter cores all the way down}

\author[J. I. Read; O. Agertz; M. L. M. Collins]{J. I. Read$^{1}$\thanks{E-mail: justin.inglis.read@gmail.com}, O. Agertz$^1$, M. L. M. Collins$^{1,2}$\\
$^1${\small Department of Physics, University of Surrey, Guildford, GU2 7XH, Surrey, UK}\\
$^2${\small Astronomy Department, Yale University, New Haven, CT 06510, USA, Hubble Fellow}\\
}

\begin{document}

\maketitle

\begin{abstract}
We use high resolution simulations of isolated dwarf galaxies to study the physics of dark matter cusp-core transformations at the edge of galaxy formation: $M_{\rm 200} = 10^7 - 10^9$\,M$_\odot$. We work at a resolution ($\sim$4\,pc minimum cell size; $\sim$250\,M$_\odot$ per particle) at which the impact from individual supernovae explosions can be resolved, becoming insensitive to even large changes in our numerical `sub-grid' parameters. We find that our dwarf galaxies give a remarkable match to the stellar light profile; star formation history; metallicity distribution function; and star/gas kinematics of isolated dwarf irregular galaxies. Our key result is that dark matter cores of size comparable to the stellar half mass radius $r_{1/2}$ {\it always form} if star formation proceeds for long enough. Cores fully form in less than 4\,Gyrs for the $M_{\rm 200} =10^8$\,M$_\odot$ and $\sim 14$\,Gyrs for the $10^9$\,M$_\odot$ dwarf. We provide a convenient two parameter `\coreNFW' fitting function that captures this dark matter core growth as a function of star formation time and the projected stellar half mass radius.

Our results have several implications: (i) we make a strong prediction that if $\Lambda$CDM is correct, then `pristine' dark matter cusps will be found either in systems that have truncated star formation and/or at radii $r > r_{1/2}$; (ii) complete core formation lowers the projected velocity dispersion at $r_{1/2}$ by a factor $\sim 2$, which is sufficient to fully explain the `too big to fail problem'; and (iii) cored dwarfs will be much more susceptible to tides, leading to a dramatic scouring of the subhalo mass function inside galaxies and groups.
\end{abstract}


\section{Introduction}\label{sec:introduction}
Dwarf galaxies provide a unique test-bed for galaxy formation and dark matter physics \citep[e.g.][]{2012AJ....144....4M}. The very smallest -- the dwarf spheroidal (dSph) galaxies of the Local Group -- are resolved in exquisite detail, providing detailed star formation histories \citep[e.g.][]{2014ApJ...789..148W}; orbits \citep[e.g.][]{2010arXiv1001.1731L}; masses \citep[e.g.][]{2009ApJ...704.1274W,2010MNRAS.406.1220W}; and dark matter mass profiles \citep[e.g.][]{2003ApJ...588L..21K,Goerdt:2006rw,2012MNRAS.426..601C,2008ApJ...681L..13B,2011ApJ...742...20W,2013NewAR..57...52B}. In particular, the shallow potential wells of the dSphs are particularly sensitive to energetic feedback from supernovae, providing unique constraints on feedback physics \citep[e.g.][]{1986ApJ...303...39D,2006MNRAS.371..885R,2013MNRAS.429.3068T}; while their high dark matter content makes them natural dark matter laboratories \citep[e.g.][]{2001ApJ...561...35D,2013MNRAS.430.2346S,2014PhRvD..89b5017H}, and excellent sites to hunt for dark matter annihilation or decay \citep[e.g.][]{1990Natur.346...39L,2011MNRAS.418.1526C,2014PhRvD..90j3506M,2015arXiv150402048B,2015arXiv150302632T,2015arXiv150302320G}.

It has been known for over a decade that tiny gas rich dwarfs favour central dark matter cores over cusps \citep{1994ApJ...427L...1F,1994Natur.370..629M}, a fact that has stood the test of time\footnote{There may be some appreciable scatter, however \citep{2005ApJ...621..757S,2014ApJ...789...63A,2014MNRAS.443.3712H}.} (e.g. \citealt{2011MNRAS.414.3617K,2011AJ....142...24O,2013MNRAS.433.2314H}). The situation for the gas-poor dSphs is more murky, with claims of cores \citep[e.g.][]{2008ApJ...681L..13B,2011ApJ...742...20W,2011MNRAS.tmp.1606A,2012ApJ...754L..39A}; cusps \citep[e.g.]{2014MNRAS.441.1584R,2014arXiv1406.6079S}; and everything in between \citep[e.g.][]{2014ApJ...791L...3B,2013ApJ...775L..30J}. However, the brightest Milky Way dSph, Fornax, is almost certainly cored\footnote{For Fornax, arguably the most robust constraint comes from its globular cluster (GC) system. Its GCs should fall to the centre of the dwarf in less than a Hubble time, creating a nucleated dwarf that is inconsistent with Fornax's observed light profile. This can be avoided if there is a central constant density core that causes dynamical friction to stall \citep{Goerdt:2006rw,2006astro.ph..6636R}. Such a timing argument has been shown to be remarkably difficult to circumvent; it holds even if Fornax is triaxial or affected by tides \citep{2012MNRAS.426..601C,2013MNRAS.431.2796K}.} \citep{2012MNRAS.426..601C}.

Such dark matter cores are not predicted by pure collisionless cold dark matter structure formation simulations \citep[e.g.][]{1991ApJ...378..496D,1996ApJ...462..563N,2008MNRAS.391.1685S,2009MNRAS.398L..21S}, which has led many to claim that this is evidence for new physics\footnote{Note that `warm dark matter' (WDM) models are often invoked as an explanation for these cores. However, WDM models cannot produce cores of the size required without fine-tuning \citep{2007PhRvD..75f1303S,2011JCAP...03..024V,2012MNRAS.424.1105M}.} \citep[e.g.][]{1994Natur.370..629M,2013MNRAS.430...81R,2014arXiv1412.1477E}. However, dark matter cores can also arise due to energetic feedback from supernovae, as first suggested by \citet{1996MNRAS.283L..72N}. \citet{2002MNRAS.333..299G} showed that the effect for a single burst is very small, suggesting that stellar feedback cannot significantly alter a dark matter cusp. However, \citet{2005MNRAS.356..107R} showed that if star formation proceeds in repeated bursts then the effect accumulates, gradually grinding a cusp down into a core. This mechanism has now been observed in hydrodynamic simulations that resolve the most massive sites of star formation (e.g. \citealt{2008Sci...319..174M,GovernatoEtAl2010,2013MNRAS.429.3068T,2014MNRAS.441.2986D,2015MNRAS.446.1140T,2015arXiv150202036O}; and for a review see \citealt{2014Natur.506..171P}), while the physics of such cusp-core transformations is now well understood\footnotemark\ \citep{2012MNRAS.421.3464P,2015arXiv150207356P}.

\footnotetext{An alternative mechanism that may act in tandem with collisionless heating from supernovae is angular momentum transfer from dense infalling lumps \citep{2001ApJ...560..636E,2010ApJ...725.1707G,2011MNRAS.416.1118C,2015MNRAS.446.1820N}. Since many cored dwarfs are observed to be bulgeless, such dense lumps must then be removed from the centre. Stellar feedback can achieve this if the dense lumps are purely gaseous \citep[e.g.][]{2015MNRAS.446.1820N}.}

Cusp-core transformations may also be key to solving two other small scale puzzles: the {\it missing satellites} problem \citep{1999ApJ...524L..19M,1999ApJ...522...82K}; and the {\it `too-big-to-fail'} problem \citep{2006MNRAS.tmp..153R,2011MNRAS.415L..40B}. The former is a large discrepancy between the number of bound dark matter structures predicted in $\Lambda$CDM and the number of satellites observed in the Local Group of galaxies. The latter is an inconsistency between the abundance and mass of satellite galaxies in $\Lambda$CDM and the Local Group. It has long been known that the missing satellites problem cannot be solved by simply pushing stars into the most massive dark matter satellites. This results in central stellar velocity dispersions that are too high to be consistent with the observed dwarf population \citep{2006MNRAS.tmp..153R}. \citet{2011MNRAS.415L..40B} showed that this problem remains even for more complex mappings from light to dark -- whether placing the dwarfs in the most massive halos before reionisation, or the most massive before infall onto the Milky Way. They called this the `too-big-to-fail' problem\footnote{This refers to the fact that it is odd that such massive dwarfs have apparently failed to form stars.}.

Several authors have noted that if dark matter halos are cored rather than cusped then this can naturally alleviate both the too-big-to-fail and missing satellites problems. Cored dwarfs have a lower central stellar velocity dispersion that -- if unmodelled -- will cause halo masses to be underestimated \citep{2006MNRAS.tmp..153R,2014ApJ...789L..17M,2014MNRAS.441.2986D,2015MNRAS.446.2363O}. Such cored dwarfs are also much more efficiently tidally stripped, which can significantly alleviate the missing satellites problem \citep{2006MNRAS.tmp..153R,2010MNRAS.406.1290P,2012ApJ...761...71Z,2013ApJ...765...22B}.

While it is becoming increasing likely that cusp-core transformations driven by bursty star formation occur in nature \citep{2012ApJ...750...33L,2013MNRAS.429.3068T,2012ApJ...744...44W,2014MNRAS.441.2717K,2015MNRAS.450.3886M}, there has been quite some debate about the mass scale at which such processes become inefficient. \citet{2012ApJ...759L..42P} use simple energetics arguments to point out that below some critical stellar mass, there simply will not be enough integrated supernovae energy to unbind a dark matter cusp. \citet{2013MNRAS.433.3539G} argue that this mass scale is sufficiently large that the cusp-core and too-big-to-fail problems cannot be solved by stellar feedback. However, \citet{2014ApJ...789L..17M} point out that it is easier to unbind cusps at high redshift, while \citet{2015arXiv150500825M} show that the assumed radial dark matter profiles matter. They find that just a few percent of the supernovae energy is sufficient to unbind cusps, even in dwarf galaxies like Fornax (see also \citealt{2014ApJ...782L..39A} and \citealt{2015arXiv150703995B}). Such calculations rely on a wide array of assumptions: the mean stellar mass to halo mass ratio; the size of the core; the coupling efficiency of the supernovae; the initial mass function of stars; the redshift of cusp-core transformations; and the dark matter radial profile, amongst others. Differences in these assumptions are likely responsible for the wide range of results reported in the literature; we will discuss this further in \ref{sec:energy}.

Alongside such energy arguments, several groups have used hydrodynamical simulations to study cusp-core transformations. These avoid some of the assumptions necessary in the above energy arguments, but nonetheless there is significant scatter between groups. \citet{2014MNRAS.441.2986D} and \citet{2015arXiv150703590T} report a peak in core formation at $\sim 10^{10}$\,M$_\odot$, with core formation ceasing below this mass scale; while \citet{2008Sci...319..174M}, \citet{2014ApJ...789L..17M}, \citet{2015arXiv150202036O} and \citet{2015arXiv151101484V} find substantial cores even in $10^9$\,M$_\odot$ dwarfs. Finally, \citet{2015MNRAS.449L..90L} have recently argued that cusps could reform as a result of late minor mergers. All of this leaves open the question of where, if anywhere, cusp-core transformations really cease. 

In this paper, we simulate isolated dwarf galaxies over the mass range $M_{200} = 10^7 - 10^9$\,M$_\odot$ to model cusp-core transformations at the very edge of galaxy formation. Such low mass dwarfs have been modelled at high redshift, typically with a view to understanding the formation of the `first galaxies' \citep[e.g.][]{2003ApJ...592..645Y,2006MNRAS.371..885R,2007ApJ...665..899W,2009arXiv0908.1254B,2009MNRAS.399...37J,2012ApJ...745...50W}. However, to date no simulations have simultaneously resolved stellar feedback and followed the evolution of such dwarfs over a Hubble time. This is the goal of this present work. All simulations presented in this paper reach a minimum cell size of $\sim 4$\,pc, allowing us to resolve the cooling occurring in the majority of {\it individual supernovae remnants}. This allows us to correctly capture the momentum injection into the interstellar medium (ISM) generated by expanding bubbles of shock heated gas. As we will show, this makes us insensitive to the details of our `sub-grid' numerical parameters (for a detailed discussion on this, see e.g. \citealt{2015arXiv150105655K}; for earlier work on the utility of resolving the interstellar medium (ISM) see e.g. \citealt{2003ApJ...590L...1K,2008PASJ...60..667S,2011MNRAS.417..950H,2013MNRAS.432.2647H,2013ApJ...770...25A}). For this reason, we consider the simulations presented here as {\it predictive}; they allow us to calculate how dark matter density profiles evolve in the face of stellar feedback, from first principles. Although we miss some potentially important physics -- mergers; tides; ram pressure stripping and photoionisation -- we show that our simulated dwarfs give a remarkable match to all known observational data for real isolated dwarfs in the Universe. This suggests that our simulations already capture the essential physics -- at least for isolated dwarfs.

This paper is organised as follows. In \S\ref{sec:simulations}, we describe the simulation suite. In \S\ref{sec:results}, we present our key results. We give an overview of the simulations (\S\ref{sec:overview}); compare them with a wide array of data for two isolated dwarfs -- Leo T and Aquarius (\S\ref{sec:observations}); and present our results for dark matter cusp-core transformations (\S\ref{sec:cuspcore}). We show that at the resolution we adopt here, our results are insensitive to our choice of `sub-grid' numerical parameters (\S\ref{sec:sensitivity}) and only weakly sensitive to our choice of initial conditions (\S\ref{sec:sensitivityICs}). We derive a convenient \coreNFW\ fitting formula that captures the evolution of the dark matter cusp as a function of the projected stellar half mass radius, and the star formation time (\S\ref{sec:coreNFW}). Finally, in \S\ref{sec:discussion} and \S\ref{sec:conclusions} we discuss our results and their implications for cosmology, and present our key conclusions.

\section{The simulations}\label{sec:simulations}

\subsection{Initial conditions}\label{sec:ics}

We set up the dark matter halos following \citet{2006MNRAS.tmp..153R}. The particles were populated using accept/reject from an analytic density profile; their velocities were drawn from a numerically calculated distribution function, assuming an isotropic velocity dispersion tensor. We assume a \citet{1996ApJ...462..563N} (hereafter NFW) form: 

\begin{equation} 
\rho_{\rm NFW}(r) = \rho_0 \left(\frac{r}{r_s}\right)^{-1}\left(1 + \frac{r}{r_s}\right)^{-2}
\label{eqn:rhoNFW}
\end{equation}
where the central density $\rho_0$ and scale length $r_s$ are given by: 
\begin{equation} 
\rho_0 = \rho_{\rm crit} \Delta c^3 g_c / 3 \,\,\,\, ; \,\,\,\, r_s = r_{200} / c
\end{equation}
\begin{equation}
g_c = \frac{1}{{\rm log}\left(1+c\right)-\frac{c}{1+c}}
\end{equation}
and
\begin{equation} 
r_{200} = \left[\frac{3}{4} M_{200} \frac{1}{\pi \Delta \rho_{\rm crit}}\right]^{1/3}
\label{eqn:r200}
\end{equation} 
where $c$ is the dimensionless {\it concentration parameter}; $\Delta = 200$ is the over-density parameter; $\rho_{\rm crit} = 136.05$\,M$_\odot$\,kpc$^{-3}$ is the critical density of the Universe at redshift $z=0$; $r_{200}$ is the `virial' radius at which the mean enclosed density is $\Delta \times \rho_{\rm crit}$; and $M_{200}$ is the `virial' mass -- the mass within $r_{200}$.

The enclosed mass for the NFW profile is given by:

\begin{equation} 
M_{\rm NFW}(r) = M_{200} g_c \left[\ln\left(1+\frac{r}{r_s}\right) - \frac{r}{r_s}\left(1 + \frac{r}{r_s}\right)^{-1}\right]
\label{eqn:MNFW}
\end{equation}
which logarithmically diverges as $r\rightarrow \infty$. We truncate the mass distribution at $r= 3\,r_{200}$. 
 
We consider four halos with virial masses $M_{200} = 10^7, 10^8, 5\times 10^8$ and $10^9$\,M$_\odot$; these have {\it dark matter} virial masses of $\Mtwodm = M_{200} / (1-f_b)$, where $f_b = 0.15$ is the Universal baryon fraction \citep[e.g.][]{2013arXiv1303.5076P}. Their concentration parameters were selected to be the cosmic mean value taken from cosmological simulations, following \citet{2007MNRAS.378...55M}. 

These halos were then filled with the Universal baryon fraction in gas ($f_b = 0.15$), set up either as an NFW profile in hydrostatic equilibrium or as a constant density slab in hydrostatic equilibrium (in order to explore our sensitivity to this choice). The gas was given a seed metallicity of $Z_{\rm gas}=10^{-3}\,Z_\odot$, representing Pop III enrichment \citep[e.g.][]{2001ApJ...548...19N,2009arXiv0908.1254B,2013RvMP...85..809K}. Experiments with the initial $Z_{\rm gas}$ in the range $10^{-10}<Z_{\rm gas}/Z_\odot<10^{-2}$ show no significant differences after the first $\sim 2$\,Gyr, as self-enrichment after the first burst of star formation becomes the dominant source of metals. Note that such differences may matter more in the presence of the UV background; we will consider this in a forthcoming paper.

We added angular momentum to the gas assuming a specific angular momentum profile as in \citet{2001ApJ...555..240B}: 

\begin{equation} 
j(r) \simeq j_{\rm max} \frac{M_{\rm NFW}(<r)}{M_{200}}
\end{equation}
where the peak specific angular momentum $j_{\rm max}$ is set such that the total halo angular momentum is given by: 

\begin{eqnarray} 
J_{\rm 200} & = & 4\pi \int_0^{\infty} j(r) \rho_{\rm NFW}(r) r^2 dr \\
& = & \lambda' \sqrt{2 G M_{200}^3 R_{\rm 200}}
\end{eqnarray}
where $\lambda'$ is the {\it spin parameter}. We assume the cosmic mean value $\lambda' = 0.035$, except for one simulation where we explore the effect of higher $\lambda' = 0.07$.

The default particle number was chosen such that each dark matter particle had mass $m_{\rm DM} = 250$\,M$_\odot$, similar to the stellar masses (see \S\ref{sec:subgrid}); one simulation was run at ten times this resolution to test for numerical convergence. We explicitly checked that our initial conditions remain in equilibrium for a Hubble time (see Appendix \ref{sec:ictest}). All simulation parameters and our simulation labelling system are summarised in Table \ref{tab:sims}. 

The above choice of `cooling halo' initial conditions has been widely used in the literature to study galaxy formation \citep[e.g.][]{2006MNRAS.370.1612K,2013MNRAS.429.3068T,2015MNRAS.452.3593H}. It has the advantage that it is well-defined and therefore useful for controlled numerical experiments. It is also likely to be realistic for tiny dwarf galaxies that undergo few mergers over most of their lifetimes \citep[e.g.][]{2015arXiv151101484V}. However, such initial conditions have the disadvantage that they start out with a dense central concentration of gas that leads unavoidably to an initial starburst \citep[see e.g. discussion in][]{2015MNRAS.452.3593H}. To explore the effect of this on our results, we ran an additional simulation: M5e8c25\_2e6\_rhocon. This is identical to our fiducial $M_{200} = 5\times 10^8$\,M$_\odot$ simulation (M5e8c25\_2e6), but starting out instead with a constant gas density out to the virial radius. We discuss the results of this simulation in \S\ref{sec:sensitivityICs}.

\begin{table*}
\resizebox{\textwidth}{!}{
\begin{tabular}{lcccccccc}
{\bf Label} & $\Mtwodm$ & $c$ & $N(<r_{\rm 200})$ & $\lambda'$ & $f_b$ & $J_{\rm 200}$ & $j_{\rm max}$ & {\bf Description}\\
& [M\,$_\odot$] & & & &  $[2.33\times 10^5 M_\odot\,{\rm kpc}\,{\rm km/s}]$ & $[{\rm kpc\,km/s}]$ &   \\
\hline
\hline
M7c36\_4e4 & $10^7\,(1-f_b)$ & $36.73$ & $4\times 10^4$ & 0.035 & 0.15 & 29.47 & 1.35 & Fiducial \\
M8c28\_4e5 & $10^8\,(1-f_b)$ & $28.57$  & $4\times 10^5$ & 0.035 & 0.15 & 1367.88 & 6.28 & Fiducial \\
M8c28\_4e5\_e001 & $10^8\,(1-f_b)$ & $28.57$  & $4\times 10^5$ & 0.035 & 0.15 & 1367.88 & 6.28 & $\epsilon_{\rm ff}=1\%$\\
M8c28\_4e5\_pST & $10^8\,(1-f_b)$ & $28.57$  & $4\times 10^5$ & 0.035 & 0.15 & 1367.88 & 6.28 & Sedov-Taylor momentum ($p_{\rm ST}$) \\
M8c28\_4e6\_r1e3 & $10^8\,(1-f_b)$ & $28.57$  & $4\times 10^6$ & 0.035 & 0.15 & 1367.88 & 6.28 & Higher DM resolution; $\rho_* = 10^3\,m_{\rm H}\,{\rm cm}^{-3}$ \\
M5e8c25\_2e6 & $5\times 10^8\,(1-f_b)$ & $24.93 $  & $2\times 10^6$ & 0.035 & 0.15 & 19996.3 & 18.38 & Fiducial \\ 
M5e8c25\_2e6\_rhocon & $5\times 10^8\,(1-f_b)$ & $24.93 $  & $2\times 10^6$ & 0.035 & 0.15 & 19996.3 & 18.38 & Constant initial gas density over $0 < r < r_{200}$ \\ 
M9c22\_4e6 & $10^9\,(1-f_b)$ & $22.23$ & $4\times 10^6$ & 0.035 & 0.15 & 63481.41 & 29.17 & Fiducial \\
M9c22\_4e6\_lam007 & $10^9\,(1-f_b)$ & $22.23$ & $4\times 10^6$ & 0.07 & 0.15 & 126962.81 & 58.34 & Double fiducial halo spin ($2\times\lambda'$)\\
\hline
\hline
\end{tabular} 
}
\caption{The simulation suite. The columns show from left to right: the simulation label; the initial NFW halo parameters (see equation \ref{eqn:rhoNFW}); the dark matter particle number $N$ within the virial radius $r_{200}$; the spin parameter $\lambda'$; the baryon fraction $f_b$; the total angular momentum $J_{200}$; the maximum specific angular momentum $j_{\rm max}$; and a description of the simulation. For further details, see \S\ref{sec:simulations}.}
\label{tab:sims}
\end{table*}

\subsection{Sub-grid physics}\label{sec:subgrid}
Our suite of simulations were carried out using the Adaptive Mesh Refinement (AMR) code {\small RAMSES} \citep{teyssier02}. The fluid dynamics is calculated using a second-order unsplit Godunov method, while the collisionless dynamics of stellar and dark matter particles is evolved using the particle-mesh technique \citep{Hockney1981}, with gravitational accelerations computed from the gravitational potential on the mesh. The potential is calculated by solving the Poisson equation using the multi-grid method \citep{GuilletTeyssier2011} for all refinement levels. The equation of state of the fluid is that of an ideal mono-atomic gas with an adiabatic index $\gamma=5/3$. 

Our refinement strategy is based on a quasi-Lagrangian approach; each cell is refined if it contains more than 8 dark matter particles, or if its baryonic mass (including gas and star particle mass) exceeds $8\times m_{\rm res}$, where $m_{\rm res}=60\Msun$ is the adopted mass resolution of the simulation. This allows the local force softening to closely match the local mean interparticle separation, which suppress discreteness effects \citep[e.g.][]{Romeo08}. Refinement is performed recursively, on a cell-by-cell basis, until the adopted maximum allowed level of refinement is reached. In our current setup, the finest grid cell size is $\Delta x\approx 4\pc$.

\subsubsection{Star formation}\label{sec:SF}
The adopted star formation and feedback physics are presented in detail in \cite{2013ApJ...770...25A} and \cite{2015ApJ...804...18A}. Briefly, we adopt a local star formation rate using a Schmidt relation
\begin{equation}
\label{eq:schmidt}
\dot{\rho}_{*}=\epsilon_{\rm ff}\frac{ \rho_{\rm g}}{t_{\rm ff}},  \ \rho>\rho_*,
\end{equation}
where $\rho_{\rm g}$ is the gas density in a cell, $t_{\rm ff}=\sqrt{3\pi/32G\rho_{\rm g}}$ the local free-fall time of the gas, and $\epsilon_{\rm ff}$ is the star formation efficiency per free-fall time. Eq.\,\ref{eq:schmidt} is sampled via a Poisson process with star particles masses being multiples of the sampling mass $m_* = 300\Msun$ \citep[see e.g.][]{dubois08}. Each star particle then looses mass due to supernovae explosions and stellar winds, see \ref{sec:FB}.

Star formation is known to correlate well with the presence of molecular gas \citep[e.g.][]{bigiel2008}, i.e. $\HH$ traced observationally by CO. We note that star formation from $\HH$ in low metallicity environments ($Z_{\rm gas}\lesssim10^{-2}\,Z_\odot$) is poorly understood, and difficult to model, due to uncertainties in dissociating UV radiation; self-shielding and shielding of $\HH$ by dust  \citep[][]{Krumholz09,Gnedin09,GnedinKravtsov11}; uncertain dust-to-gas ratios \citep{Fisher2014}; and the physics of line overlap \citep{GnedinDraine2014}. For these reasons, we adopt here a high star formation density threshold, $\rho_*=300\,m_{\rm H}\,{\rm cm}^{-3}$, ensuring that star formation proceeds only in gas dense enough to be mainly molecular gas. We explore the effect of raising this to $\rho_* = 10^3\,m_{\rm H}\,{\rm cm}^{-3}$ in one run.

In our fiducial models, we adopt a star formation efficiency per free fall time $\epsilon_{\rm ff} =0.1$ motivated by observations of local GMCs \citep{lada_etal10,Murray2011b,evans_etal14}, as well as recent numerical work \citep[][]{Hopkins2014,2015ApJ...804...18A} demonstrating how the interplay between local efficient star formation and stellar feedback reproduces the low {\it global} star formation efficiency inferred from the Kennicutt-Schmidt relation \citep[e.g.][]{bigiel2008} and global galaxy scaling relations. In one run, we explore the effect of lowering this by a factor of ten.

\subsubsection{Stellar Feedback}\label{sec:FB}
We adopt the stellar feedback model described in \cite{2013ApJ...770...25A}. Briefly, each formed stellar particle is treated as a single-age stellar population with a \cite{chabrier03} initial mass function (IMF). We account for injection of energy, momentum, mass and heavy elements over time via supernova type II (SNII) and type Ia (SNIa) explosions, stellar winds and radiation pressure (allowing for both single scattering and multiple scattering events on dust) on the surrounding gas. Each mechanism depends on the stellar age, mass and gas/stellar metallicity, calibrated on the stellar evolution code {\small STARBURST99} \citep{Leitherer1999}. Feedback occurs continuously at the appropriate times when the various feedback process are known to operate, taking into account the lifetime of stars of different masses in a stellar population. To track the lifetimes of stars within the population we adopt the metallicity dependent age-mass relation of \cite{Raiteri1996}. 

Momentum from stellar winds, radiation pressure, and SNe blastwaves is added to the 26 nearest cells surrounding a parent cell of the stellar particle. Thermal energy from shocked SNe and stellar wind ejecta is injected directly into the parent cell. In the case of SNe we model these as discrete events (i.e. we sample the stellar IMF discretely) and add $10^{51}$ ergs per SN to the gas thermal energy. As the average gas and stellar metallicities in our suite of simulated dwarf galaxies are low ($Z_{\rm gas}<0.1$, see \S\,\ref{sect:zdist}), we expect the impact from stellar winds and radiation pressure on dust to be relatively small compared to that of the SNe. Heavy elements (metals) injected by supernovae and winds are advected as a passive scalar and are incorporated self-consistently in the cooling and heating routine. The code accounts for metallicity dependent cooling by using tabulated cooling functions of \cite{sutherlanddopita93} for gas temperatures $10^4-10^{8.5}\,$K, and metal fine-structure cooling below $10^4$ K as in \cite{rosenbregman95}. 

At the current numerical resolution ($\Delta x\sim 4\pc$), and for the density field and typical metallicities in these simulations, we are likely to resolve the impact of most SNe explosions. For a point like explosion, the cooling radius, i.e. the  bubble radius for which radiative losses are expected to be important for each discrete SN event, scales as $r_{\rm cool}\approx 30 n_0^{-0.43}(Z_{\rm gas}/Z_\odot+0.01)^{-0.18}$ pc for a supernova explosion with energy $E_{\rm SN}=10^{51}$ erg and ambient gas density $n_0$ (in units of $\cc$; e.g. \citealt{Cioffi1988,Thornton1998,KimOstriker2014}). \cite{KimOstriker2014} \citep[and see also][]{Martizzi2014,Gatto2014,Simpson2014} demonstrated that if $r_{\rm cool}$ is resolved by at least three grid cells ($r_{\rm cool}\geq3\Delta x$), the correct amount of momentum generated in the Sedov-Taylor phase is recovered before the bubble energy is radiated away. 

For typical ISM gas densities, this momentum, $p_{\rm ST}\approx 2.6\times 10^5\,E_{51}^{16/17}n_0^{-2/17} \Msun\kmsec$ \cite[e.g.][]{Blondin1998}, can be $\gtrsim 10$ times greater than the initial ejecta momentum, which is why it is important to capture this process. For $\Delta x=4\pc$, and $Z_{\rm gas}=10^{-2}\,Z_\odot$, we resolve single explosions up to ambient densities of $\approx 50\cc$. Type II SNe occur for almost $\sim 40\Myr$ (roughly the age of an $8\Msun$ star) in our models, and during this time pre-SNe feedback and gas turbulence preprocesses the gas, allowing for explosions to be resolved. In \S\ref{sec:sensitivity} we demonstrate that this is the case, and that our conclusions  are insensitive to wether we explicitly inject $p_{\rm ST}$, rather than allowing it to be generated ``above the grid" via individual expanding SNe bubbles.

\begin{figure*}
\begin{center}
\includegraphics[width=0.95\textwidth]{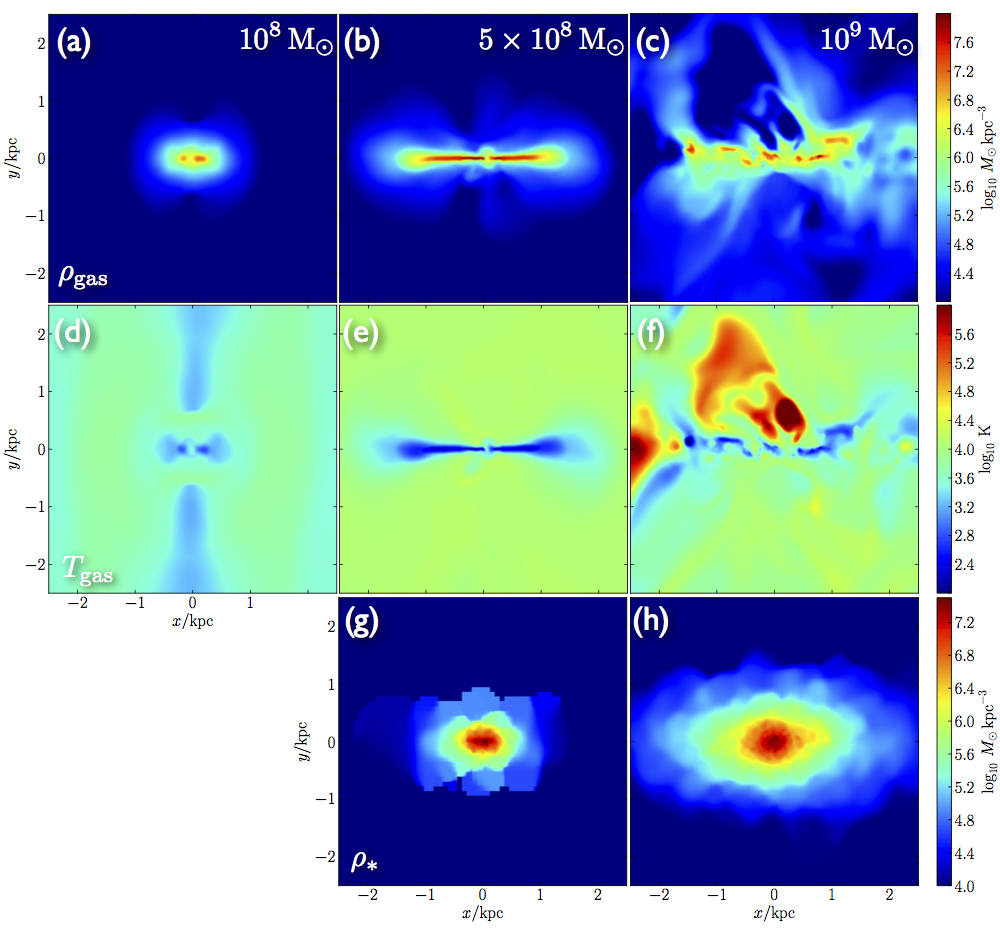}
\caption{Overview of the simulation results for our fiducial simulations: {\bf (a-c)} projected gas density $\rho_{\rm gas}$; {\bf (d-f)} gas temperature $T_{\rm gas}$; and {\bf (g-h)} stellar density $\rho_*$, for an edge on view of the galaxies. The columns show the $M_{200} = 10^8$\,M$_\odot$, $5 \times 10^8$\,M$_\odot$ and $10^9$\,M$_\odot$ fiducial simulations (M8c28\_4e5; M5e8c25\_2e6; M9c22\_4e6; see Table \ref{tab:sims}), respectively. Notice that very few stars form in the $10^8$ simulation -- just 86 star particles -- which is too few to make a plot of $\rho_*$. By contrast, the $10^9$\,M$_\odot$ simulation forms over 15,000 star particles, and overlapping SNe explosions drive large hot bubbles in the disc as can be seen in panels (c) and (f). We do not show results for the $10^7$\,M$_\odot$ simulation since this forms just 1-2 star particles after which star formation ceases.}
\label{fig:overview} 
\end{center}
\end{figure*}

\section{Results}\label{sec:results}

\subsection{Overview}\label{sec:overview}

In Figure \ref{fig:overview}, we give an overview of our simulation results over the mass range $10^8 - 10^9$\,M$_\odot$. From top to bottom, the rows show the projected gas density $\rho_{\rm gas}$; temperature $T_{\rm gas}$; and the stellar density $\rho_*$ for an edge on view of the galaxies. The columns show the $10^8$\,M$_\odot$, $5 \times 10^8$\,M$_\odot$ and $10^9$\,M$_\odot$ simulation, respectively. Notice that very few stars form in the $10^8$ simulation -- just 86 star particles -- which is too few to make a plot of $\rho_*$. By contrast, the $10^9$\,M$_\odot$ simulation forms over 15,000 star particles, and overlapping SNe explosions drive large hot bubbles in the disc as can be seen in panels (c) and (f).

We do not show results for the $10^7$\,M$_\odot$ simulation since this forms just 1-2 star particles after which star formation ceases. This agrees with earlier works that suggest that stellar feedback sets an edge to galaxy formation at $M_{200} \sim 10^7$\,M$_\odot$ \citep[e.g.][]{2006MNRAS.371..885R}. However, unlike these previous works, here galactic winds are not modelled in a sub-grid fashion, but emerge self-consistently by resolving individual interacting SNe explosions.

We note that the precise edge of galaxy formation will be very sensitive to the details of reionisation; epoch of gas infall; star formation; cooling physics; and `population III' star formation/evolution that we do not properly resolve or model here \citep[e.g.][]{Efstathiou1992,Ricotti2008,2009MNRAS.399...37J,2009arXiv0908.1254B,2012ApJ...745...50W}. For this reason, while we are confident that galaxies at $10^7$\,M$_\odot$ will contain very few stars, we defer further exploration of this mass scale to future work where we will consider such physics more carefully. 

\subsection{Observational properties}\label{sec:observations} 

Before discussing the evolution of the dark matter distribution in our simulations, we first compare them with a host of observational data for two isolated field dwarfs: Leo T and Aquarius (also called DDO 210). These are two of the smallest galaxies in the Universe with on-going star formation \citep{2007ApJ...656L..13I,2008MNRAS.384..535R,1959PDDO....2..147V}. At $417^{+20}_{-19}$\,kpc \citep{2007arXiv0706.0516S} and $\sim 1$\,Mpc \citep{1999AJ....118..853L} from any large spiral, they are extremely isolated. Both have measured star formation histories \citep{2012ApJ...748...88W,2014ApJ...795...54C}; gas kinematics \citep{2004A&A...413..525B,2008MNRAS.384..535R,2015AJ....149..180O}; stellar kinematics \citep{2007arXiv0706.0516S,2014MNRAS.439.1015K}; photometry \citep{2008ApJ...684.1075M,2006MNRAS.373..715M}; and stellar metallicity distribution functions \citep{2012ApJ...748...88W,2013ApJ...779..102K,2014ApJ...795...54C}. For this reason, they are the natural first place to look for data comparisons with the simulations we present here. The results from the $M_{200}=5\times10^8\Msun$ and $10^9\Msun$ simulations (see Table\,\ref{tab:sims}) are compared to observational data for Leo T and Aquarius in Figures \ref{fig:sim_data_5e8} and \ref{fig:sim_data_1e9}. The panels show the stellar surface density (a); star formation rate (b); stellar metallicity distribution function (c); gas rotation curve (d); gas velocity dispersion (e); and the projected stellar velocity dispersion (f). We discuss these in turn, next. 

\subsubsection{Projected light profiles}\label{sec:projlight}

The vertical green dashed lines in Figures \ref{fig:sim_data_5e8}(a) and \ref{fig:sim_data_1e9}(a) mark the projected 2D half mass radius of the simulations; the red bands mark the similar confidence intervals for LeoT \citep{2008ApJ...684.1075M} and Aquarius \citep{2006MNRAS.373..715M}, respectively. The red and magenta data points show data from resolved star counts for LeoT \citep{2008ApJ...684.1075M} and Aquarius \citep{2008ApJ...684.1075M}; we renormalise these to match our simulated surface mass density at the half stellar mass radius. The brown data points in Figure \ref{fig:sim_data_1e9}(a) show the stellar surface mass density for Aquarius derived from integrated light \citep{2012AJ....143...47Z}. This is also renormalised to match our simulated profile at the projected half stellar mass radius\footnote{If we do not renormalise, then we find that the observed stellar surface density profile is lower than our fiducial $M_{200} = 10^9\Msun$ simulation by a factor $\sim 5-7$. However, there is at least a factor $\sim 2$ uncertainty in the stellar mass of Aquarius \citep{2013ApJ...779..102K,2012AJ....143...47Z}, while our simulations overproduce stars in the first $2-4$\,Gyrs due to our idealised `cooling halo' initial conditions (see discussion in \S\ref{sec:ics} and \S\ref{sec:sfh}). If we exclude stars from the transient start-up phase, and take the larger \citet{2013ApJ...779..102K} stellar mass for Aquarius, then our $10^9\Msun$ fiducial simulated dwarf gives an excellent match to the stellar surface mass density profile for Aquarius in both its normalisation and its shape.}.

The data in \citet{2006MNRAS.373..715M} for Aquarius are sufficiently good that the surface brightness could be split by age into old and young stars ($<60$\,Myrs; magenta data points in Figure \ref{fig:sim_data_1e9}(a)). Performing a similar cut on our simulations, we qualitatively recover the observed result that the young stars (red lines) are more concentrated than the full distribution (blue), with a steeper surface brightness fall-off. This occurs because the stars are collisionlessly heated by stellar feedback, similarly to the dark matter (see \S\ref{sec:cuspcore} and e.g. \citealt{2005MNRAS.356..107R,2012ApJ...750...33L,2013MNRAS.429.3068T,2015arXiv151201235E,2015arXiv151203538G}). Over time, this leads to the older population becoming hotter and more extended than the younger stellar population. Such heating occurs very rapidly, as shown by the red lines on panel Figure \ref{fig:sim_data_1e9}(a). These mark the stellar surface mass density for two age cuts at  $<100$ and $<200$\,Myrs. Notice that the stars $<200$\,Myrs old are already substantially more extended than those $<100$\,Myrs old.

While our $M_{200} = 10^9\Msun$ fiducial simulation gives a remarkable match to the real data for Aquarius, our $M_{200} = 5\times 10^8\Msun$ fiducial simulation does not rise as steeply as the data for Leo T (compare the red data points and solid blue line in Figure \ref{fig:sim_data_5e8}(a)). Interestingly, however, if we incline our simulation at 20$^\circ$ (i.e. near face-on), then it does give an excellent match (blue dashed line). We will return to this when discussing LeoT's rotation curve and projected stellar velocity dispersion, below.

\subsubsection{Star formation histories}\label{sec:sfh}

The star formation histories are given in Figures \ref{fig:sim_data_5e8} and \ref{fig:sim_data_1e9} panels (b). These show the star formation rate (SFR) as a function of time $t$, where $t=0$ is the {\it beginning} of the Universe. The blue lines show the raw simulation results; the red and purple show data for Leo T and Aquarius, respectively (taken from \citealt{2012ApJ...748...88W} and \citealt{2014ApJ...795...54C}); the green data points show the simulation results binned similarly to the observational data for Leo T. The bursty nature of the star formation in the simulations is evident; this is what is ultimately responsible for the dark matter cusp-core transformations that we will discuss in \S\ref{sec:cuspcore} \citep[c.f.][]{2005MNRAS.356..107R,2012MNRAS.421.3464P}. However, when binned at the age resolution of observed star formation histories (that come from fits to the colour magnitude diagram of stars), the bursts are much harder to spot. At this age resolution, the simulations -- the like observational data -- look very smooth, with a near-continuous star formation history over cosmic time.

Our $5 \times 10^8$\,M$_\odot$ simulation gives a remarkable match to the data for Leo T, while the $10^9$\,M$_\odot$ simulation gives a much better match to Aquarius. This suggests that the star formation history is very sensitive to the halo mass. Indeed, we may imagine using the star formation history, calibrated on these models, as a sensitive indicator of $M_{200}$ in low mass systems. This is potentially very powerful as it can work as a mass indicator even for dwarfs that have had their star formation truncated on in-fall to a galaxy or group. We will discuss such ideas further in forthcoming publications. 

While the mean star formation rate agrees very well between the simulations and the data, there are differences in the fine structure. Looking at Figures \ref{fig:sim_data_5e8} and \ref{fig:sim_data_1e9} panel (b), we can see that both simulations overproduce stars over the first $\sim 2-4$\,Gyrs with respect to Leo T and Aquarius (compare the green data points with the red/purple data points). This owes to our idealised initial conditions. By starting out with a fully-formed dark matter halo filled with dense gas, an initial starburst is inevitable. We discuss the sensitivity of our results to our choice of initial gas density distribution in \S\ref{sec:sensitivityICs}.

In real galaxies, early star formation will be suppressed both by reionisation and mergers, neither of which are modelled here. The former pre-heats gas, suppressing cooling and star formation; the latter builds galaxies out of smaller sub-units that form stars less efficiently. Indeed, Aquarius shows statistically significant evidence for an enhancement in star formation $\sim 6$\,Gyrs ago that could owe to some late gas accretion (possibly due to a merger; see the discussion in \citealt{2014ApJ...795...54C}). We discuss these issues further in \S\ref{sec:discussion}, but note here that such differences between the theoretical models and the data are much smaller than the difference in mean star formation rate between Aquarius and Leo T (or equivalently, between our $M_{200} = 5 \times 10^8$\,M$_\odot$ fiducial simulation and our $M_{200} = 10^9$\,M$_\odot$ simulation).

\subsubsection{Stellar metallicity distribution functions} \label{sect:zdist}

In Figures \ref{fig:sim_data_5e8}(c) and \ref{fig:sim_data_1e9}(c), we compare our simulated stellar metallicity distribution functions (MDFs; blue histograms) with data for Leo T and Aquarius (red bands). For the data, we use the mean metal content [M/H] as determined from fits to the colour magnitude diagram. This measure is closest to our simulated MDFs that are all scaled relative to solar abundance. The spectroscopic measures of [Fe/H] are, however, in both cases very similar and do not affect the conclusions we draw here \citep{2013ApJ...779..102K}. The width of the red bands shows the measured variance of [M/H]; in both cases, this is also in excellent agreement with the simulations. 

The above results indicate that galactic winds have regulated the gas metal reservoir in a realistic manner. To better understand the role of metal rich outflows, we calculate effective yield, defined as:
\begin{equation}
\label{eq:yield}
y_{\rm eff}=\frac{Z_{\rm gas}}{\ln(1/f_{\rm gas})},
\end{equation}
where $f_{\rm gas}= M_{\rm gas}/(M_{\rm gas}+M_\star)$ is the fraction of baryons in the gas phase. The effective yield has been widely used as a diagnostic of the evolution of the baryonic component of galaxies, and more specifically as a test of the validity of the closed-box approximation \citep{PagelPatchett1975,Edmunds1990}. Observationally, the effective yield is known to decrease with galactic mass \citep{Tremonti2004}, with a sharp decline around the mass of dwarf galaxies \citep[$v_{\rm rot}\lesssim 100\,{\rm km}\,{\rm s}^{-1}$,][]{Garnett02}.

Under a ``closed-box" assumption (no inflow or outflow of material), the effective yield is always equal to the true yield $y_{\rm true}$. This is typically defined for a single stellar population as the mass in newly synthesised metals returned to the ISM, normalised to the stellar mass of this population in stellar remnants and long-lived stars. For our feedback prescription, and choice of IMF, $y_{\rm true}=0.022$.

Following \cite{Tassis08}, \citep[see also][]{Brooks07,2015ApJ...804...18A}, we calculate the observed effective yield using Equation\,\ref{eq:yield}, where we consider only the cold gas ($T\le10^{4}$ K) and the metal content within the stellar extent, defined as the radius that includes 90\% of the total stellar mass. The effective yields for the LeoT and Aquarius simulations are $y_{\rm eff}\approx 4.8\times 10^{-3}$ and $5.5\times 10^{-3}$, respectively. These values are lower than the true yield, indicating that the gas and stellar metallicities (Figures \ref{fig:sim_data_5e8} and \ref{fig:sim_data_1e9}) are controlled by supernovae driven outflows.

\subsubsection{Gas kinematics}

\begin{figure*}
\begin{center}
\includegraphics[width=0.99\textwidth]{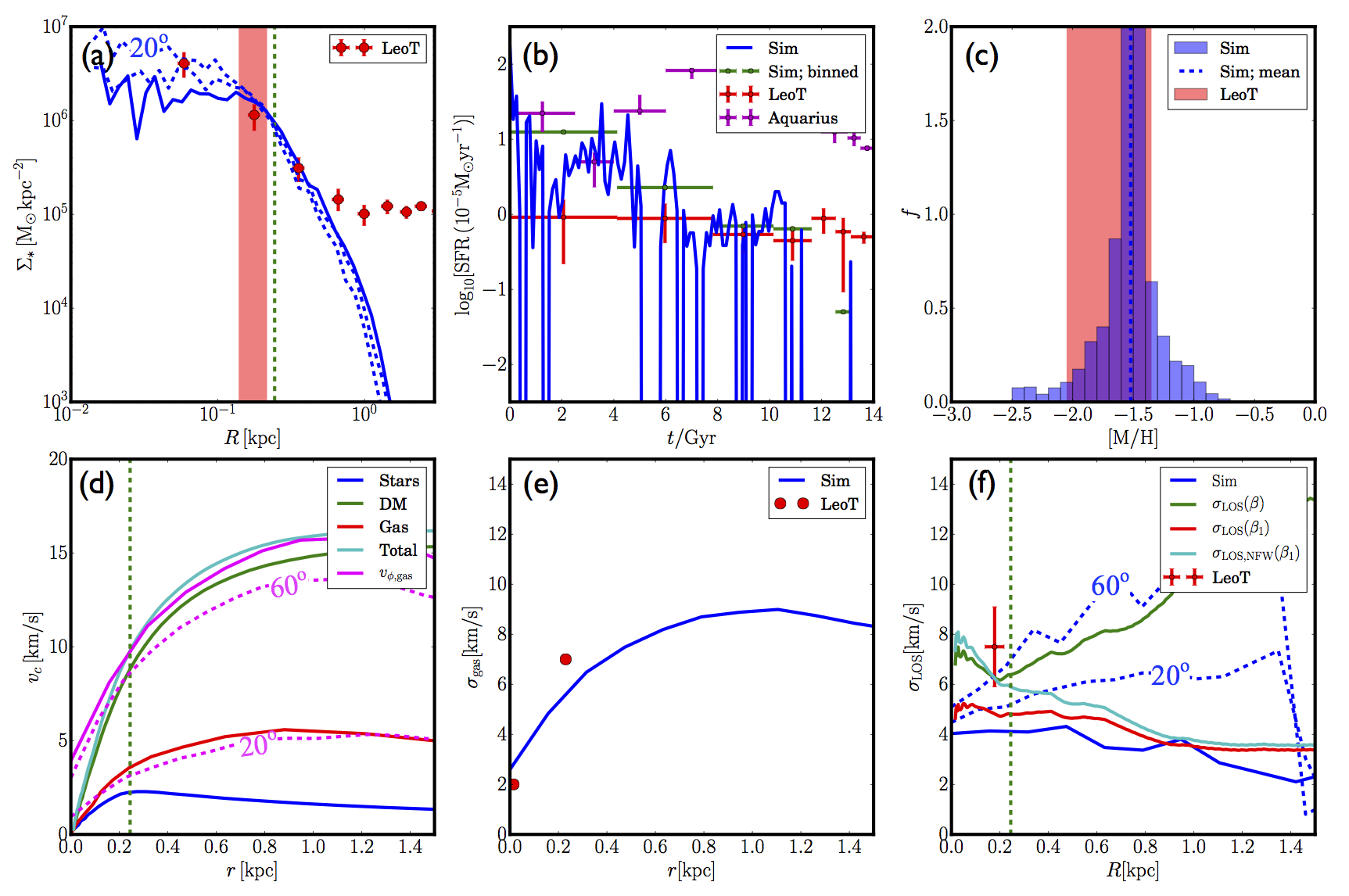}
\caption{A comparison of our $M_{\rm 200} = 5\times 10^8$\,M$_\odot$ fiducial simulation (M5e8c25\_2e6) with a host of observational data for the Leo T isolated dwarf irregular galaxy. The panels show: {\bf (a)} The stellar surface density $\Sigma_*$. The simulation results for an edge-on projection are marked by the blue lines; the dashed lines show the same inclined at 60$^\circ$ and 20$^\circ$, as marked (where $0^\circ$ corresponds to a face-on view). The vertical green dashed line marks the projected stellar half mass radius (and similarly for all other panels). The red data points show the photometric data for Leo T, taken from \citet{2008ApJ...684.1075M}; the red band marks the observed projected half mass radius, including uncertainties. {\bf (b)} The star formation rate (SFR) as a function of time $t$ (where $t=0$ is the beginning of the Universe). The blue line shows the full simulation results; the green re-binned to match the Leo T data. Overplotted are measured star formation histories for Leo T and Aquarius, taken from \citet{2012ApJ...748...88W,2014ApJ...795...54C}. {\bf (c)} Stellar metallicity distribution function, using the scaled Solar mean metallicity [M/H]. The histogram shows the simulation results; the mean is marked by the vertical dashed line. The data for Leo T are marked by the red band; the width marks the measured dispersion (data taken from \citealt{2012ApJ...748...88W}). {\bf (d)} Gas rotation velocity $v_{\phi,{\rm gas}}$ (magenta line). The magenta dashed lines mark two different inclination angles (20$^\circ$; 60$^\circ$, as marked). The blue, red, green and cyan lines mark the circular speed curves (measured directly from the gravitational potential) for the stars, gas, dark matter, and all matter, respectively. {\bf (e)} Gas velocity dispersion (blue). Two data points for Leo T (reported without errors) are overplotted, taken from \citet{2008MNRAS.384..535R}. {\bf (f)} Projected stellar velocity dispersion. The simulation data are shown in blue; the dashed lines give results for two different inclination angles (20$^\circ$; 60$^\circ$, as marked). The red line shows a Jeans equation estimate of $\sigma_{\rm LOS}$ that accounts for the significant stellar rotation (see text for details). The green line shows $\sigma_{\rm LOS}$ with this rotation removed. The cyan line shows $\sigma_{\rm LOS}$ if we restore the initial dark matter cusp, all other things being equal. Notice that the green line produces a dispersion profile similar to that seen if Leo T is nearly edge-on, while reinstating the dark matter cusp (cyan line) boosts the central $\sigma_{\rm LOS}$ by nearly a factor of two (compare the blue and cyan lines). The red data point shows the measured $\sigma_{\rm LOS}$ for Leo T \citep{2007arXiv0706.0516S}.}
\label{fig:sim_data_5e8} 
\end{center}
\end{figure*}

\begin{figure*}
\begin{center}
\includegraphics[width=0.99\textwidth]{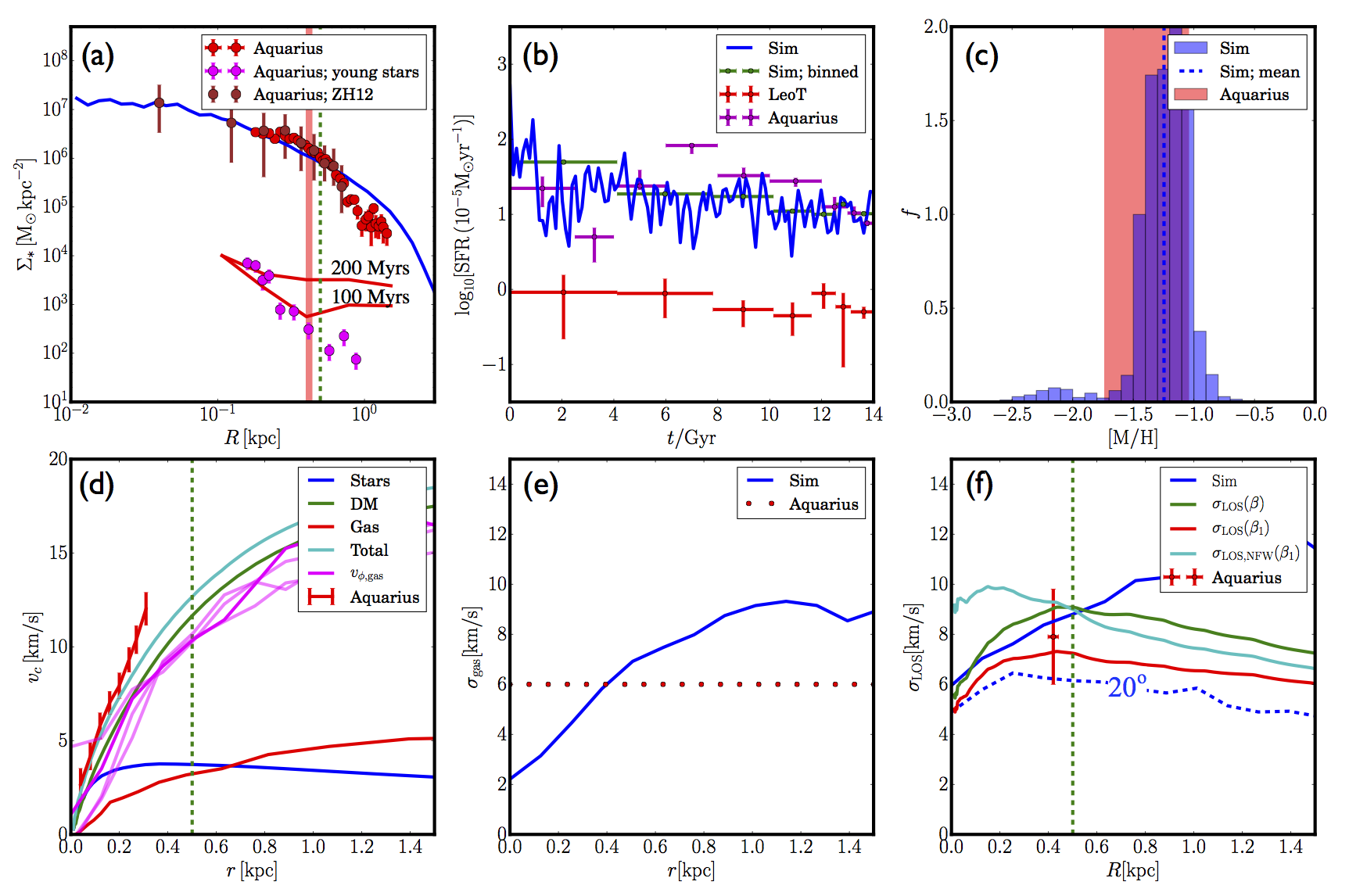}
\caption{A comparison of the $M_{\rm 200} = 10^9$\,M$_\odot$ fiducial dwarf (M9c22\_4e6) with a host of observational data for the Aquarius dwarf irregular galaxy. The panels are as in Figure \ref{fig:sim_data_5e8}, with two exceptions. In panel {\bf (a)}, we show results also for the young stars ($<100$ and $<200$\,Myrs old, as marked in red), and similarly for the Aquarius data ($<60$\,Myrs old; magenta data points). In panel {\bf (d)}, we show results for $v_{\phi,{\rm gas}}$ at different times $[0,0.5,1,1.5]$\,Gyrs ago (magenta lines). The scatter is caused by the bursty star formation that blows large HI bubbles through the disc. Data for the asymmetric drift corrected Aqarius rotation curve (red data points) are overplotted (taken from \citealt{2015AJ....149..180O}). The data for Aquarius are compiled from: \citet{2006MNRAS.373..715M,2012AJ....143...47Z} (ZH12); \citet{2014ApJ...795...54C,2015AJ....149..180O}; and \citet{2014MNRAS.439.1015K}.}
\label{fig:sim_data_1e9} 
\end{center}
\end{figure*}

Figures \ref{fig:sim_data_5e8}(d) and \ref{fig:sim_data_1e9}(c) show the rotation velocity of the gas $v_{\phi,{\rm gas}}$ (magenta lines). We plot results only for gas with temperature $T < 10^4$\,K to mimic HI gas observations. For the $M_{200} = 5 \times 10^8$\,M$_\odot$ simulation (Figure \ref{fig:sim_data_5e8}), we plot the rotation curve also at two different inclination angles, $20^\circ$ and $60^\circ$ (dashed lines, as marked; $0^\circ$ corresponds to a face on galaxy). For the $10^9$\,M$_\odot$ simulation (Figure \ref{fig:sim_data_1e9}), we show the time dependance of $v_{\phi,{\rm gas}}$. The translucent magenta lines show $v_{\phi,{\rm gas}}$ at times $[0,0.5,1,1.5]$\,Gyrs ago. As can be seen, there is appreciable scatter in $v_{\phi,{\rm gas}}$ as a function of time. We find that this scatter is slightly larger for the high spin version of this simulation, M9c22\_4e6\_lam007. However, since the results for M9c22\_4e6\_lam007 are otherwise very similar to M9c22\_4e6, we omit it for brevity. Also marked on the plots are the circular speed curves for the stars (blue); dark matter (green); gas (red); and all mass (cyan). For the $10^9$\,M$_\odot$ simulation, we overplot the asymmetric drift corrected rotation curve data for Aquarius (red data points), taken from \citet{2015AJ....149..180O}. Panel (e) shows that the gas velocity dispersions (blue), as compared to data (red). 

For Leo T (Figure \ref{fig:sim_data_5e8}), no rotation curve has been reported, but the gas velocity dispersion of the $5 \times 10^8$\,M$_\odot$ simulation is in excellent agreement with reported measurements (\citealt{2008MNRAS.384..535R}; red dots). Interestingly, if Leo T is inclined at $\sim 20^\circ$ -- as would give the best fit to the photometric light profile (see \S\ref{sec:projlight}) -- then the gas rotation would indeed be very difficult to detect, with a peak non-inclination corrected $v_{\phi,{\rm gas}}$ of just under $\sim 5$\,km/s (magenta dashed line). When aligned edge-on, the gas rotational velocity (magenta solid line) gives a reasonable match to the circular speed curve (cyan line), with almost all of the mass at all radii coming from the dark matter (solid green line). The gas and star contributions (red and blue lines) are comparable, similarly to what has been reported recently for low mass isolated dwarf irregulars in the LittleTHINGS survey \citep{2015AJ....149..180O}. We will compare our models to rotation curves in more detail in a companion paper. 

Our $10^9$\,M$_\odot$ simulation gives a good qualitative match to the observed rotation curve and gas dispersion of Aquarius (see panels (d) and (e) in Figure \ref{fig:sim_data_1e9}), if aligned edge on. (For this reason, we do not consider alternative inclination angles for our $10^9$\,M$_\odot$ simulation.) However, the amplitude of rotation is somewhat higher for Aquarius, suggesting that it is either slightly more massive than $M_{200} = 10^9$\,M$_\odot$, and/or has higher concentration.

\subsubsection{Projected stellar velocity dispersions} 

Finally, we consider the observed projected stellar velocity dispersions. For both Leo T and Aquarius, the latest data amount to a single measurement at $\sim$ the projected stellar half light radius. This is shown in Figures \ref{fig:sim_data_5e8} and \ref{fig:sim_data_1e9}, panels (f). The blue lines show the simulation results aligned edge on; the dashed aligned at $20^\circ$ and $60^\circ$, as marked. Notice that for Leo T, the $20^\circ$ inclination gives a reasonable match to the data, but is $\sim 2$\,km/s lower. The red line shows a fit to the simulation data using the Jeans equation, where we account also for the fact that the stars rotate significantly in the $x-y$ plane. Specifically, we solve the radial Jeans equation \citep[e.g.][]{1987gady.book.....B}: 

\begin{equation}
\sigma_{\rm LOS}^2 = \frac{2}{\Sigma_*(R)} \int_R^\infty \left(1 - \beta\frac{R^2}{r^2}\right)\frac{\nu_*(r)\sigma_r^2(r)r}{\sqrt{r^2-R^2}}dr
\label{eqn:sigmaLOS}
\end{equation}
where $\Sigma_*(R)$ is the surface brightness density of the stars; $\nu_*(r)$ is the spherically averaged stellar density;

\begin{equation}
\beta = 1-\frac{\sigma_\theta^2 + \sigma_\phi^2}{2\sigma_r^2}
\end{equation}
is the {\it velocity anisotropy}; and $\sigma_{r,\theta,\phi}$ are the radial, $\theta$ and $\phi$ velocity dispersions, respectively. The radial dispersion relates to the enclosed mass $M(r)$ as: 

\begin{equation} 
\sigma_r^2 = \frac{f(r)}{\nu_*} \int_r^\infty \frac{GM(r')\nu_*(r')f(r')}{r'^2}dr'
\end{equation}
where:

\begin{equation}
f(r) = \exp\left(-2\int_0^r \frac{\beta'}{r'}dr'\right)
\end{equation}
and: 
\begin{equation}
\beta' = 1-\frac{\sigma_\theta^2 + \sigma_\phi^2 + \overline{v}_\phi^2}{2\sigma_r^2}
\end{equation}
where $\beta'$ accounts for the mean streaming motion in the plane $\overline{v}_\phi$. This is an important contributor since our simulated stellar distributions are significantly rotating. This rotation explains why the face-on projected velocity dispersions are rather low (Figures \ref{fig:sim_data_5e8}(f) and \ref{fig:sim_data_1e9}(f) solid blue lines). 

We use our perfect knowledge of $\nu_*, \Sigma_*, \beta', \beta$ and $M(r)$ to determine a `Jeans equation' $\sigma_{\rm LOS}$, by numerically solving equation \ref{eqn:sigmaLOS}; this is marked by the red lines on Figures \ref{fig:sim_data_5e8}(f) and \ref{fig:sim_data_1e9}(f). For the $5 \times 10^8$\,M$_\odot$ simulation, this gives an excellent match to the face-on $20^\circ$ simulation data (blue dashed line). For the $10^9$\,M$_\odot$ simulation, it is slightly off at large radii, most likely owing to the fact that the system is not quite in a steady-state (a key assumption of the Jeans analysis). The agreement is nonetheless good. 

Armed with our Jeans analysis, we can now determine what the stellar velocity dispersion would look like in the absence of rotation or if we `undo' the cusp-core transformation. The green lines show the effect of switching off rotation; the cyan of enforcing the initial NFW profile, all other things being equal. Removing rotation boosts the projected velocity dispersion, particularly at large radii. This may be closer to the situation in real dwarfs if they undergo significant late mergers. We do not model such mergers here, but they will act to randomise the rotation, lowering $\overline{v}_\phi$ and moving us more towards the green lines. Switching to a cuspy NFW profile mostly alters the central velocity dispersions inside $\sim r_{1/2}$. The effect, however, is dramatic -- nearly doubling $\sigma_{\rm LOS}$ in both cases (compare the blue and cyan solid lines in Figures \ref{fig:sim_data_5e8} and \ref{fig:sim_data_1e9}, panels (f)). This highlights that using stellar velocity dispersions alone to estimate halo masses could lead to highly biased conclusions, if assuming an NFW dark matter mass profile. We discuss this further in \S\ref{sec:discussion}. 

We stress that {\it neither of our simulations has been fit to the data}. Nor has any `fine tuning' of our simulation parameters been performed. Our only freedoms are in the inclination angle of our simulations; and which galaxy we choose to compare our simulations to. For this reason, it is striking that the agreement of both simulations with Leo T and Aquarius, respectively, is so good. It suggests that we have captured the essential physics relevant for modelling isolated dwarfs in the field. We discuss this further in \S\ref{sec:discussion}.

\subsection{Dark matter cusp-core transformations}\label{sec:cuspcore}

\begin{figure*}
\begin{center}
\includegraphics[width=0.95\textwidth]{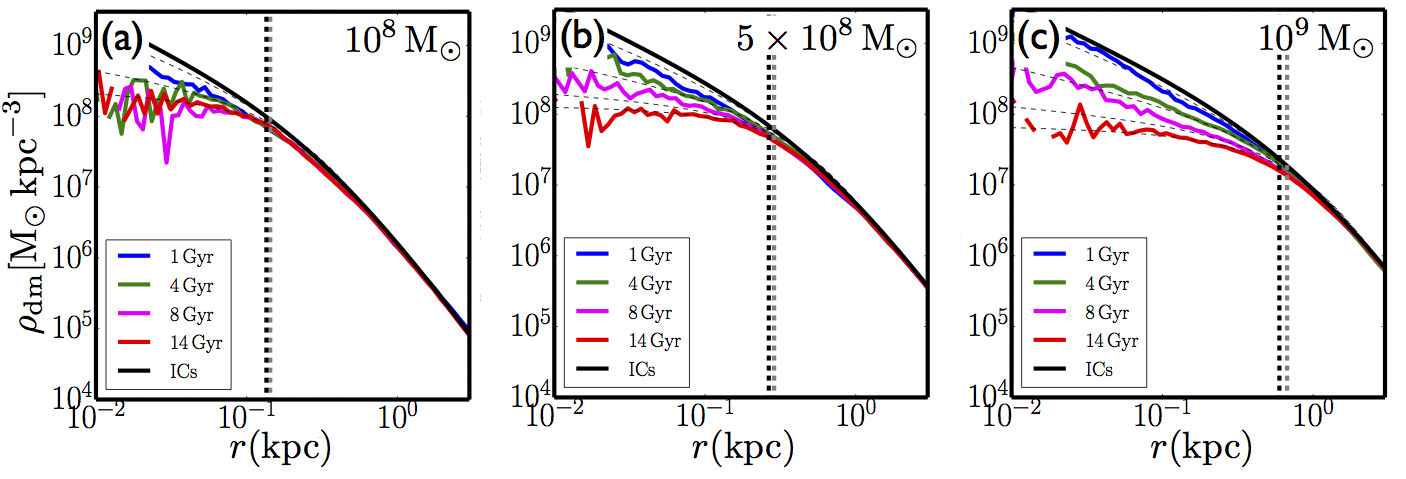}
\caption{Dark matter cusp-core transformations. From left to right, the panels show the time evolution of the dark matter halo density profiles in our $10^8$\,M$_\odot$, $5 \times 10^8$\,M$_\odot$ and $10^9$\,M$_\odot$ fiducial simulations, respectively (M8c28\_4e5; M5e8c25\_2e6; M9c22\_4e6; see Table \ref{tab:sims}). The black lines mark the initial NFW profiles; the coloured lines show the time evolution, as marked. The vertical grey and black lines mark the 3D stellar half mass radius ($r_{1/2}$) and 2D projected stellar half mass radius ($R_{1/2}$), respectively. The thin grey dashed lines show the \coreNFW\ profile fitting function (see \S\ref{sec:coreNFW} for further details).}
\label{fig:cuspcoretransforms}
\end{center}
\end{figure*}

We now turn our attention to the evolution of the underlying dark matter halos in our simulations. In Figure \ref{fig:cuspcoretransforms}, we show the time evolution of the dark matter density profile for our $M_{200} = 10^8$\,M$_\odot$ (left), $5 \times 10^8$\,M$_\odot$ (middle) and $10^9$\,M$_\odot$ (right) fiducial simulations. Marked on each panel is the 3D ($r_{1/2}$) and projected ($R_{1/2}$) stellar half mass radius of the stars (grey and black vertical dashed lines, respectively), and the time evolution of the dark matter density profile (coloured lines, as marked). The thin black dashed lines show a fitting function that we discuss in \S\ref{sec:coreNFW}. 

As can be seen, {\it in all cases the initial cusps are transformed into cores of size $\sim$ the 3D stellar half mass radius $r_{1/2}$} (grey vertical dashed lines). This should perhaps not be surprising. As demonstrated in previous works, potential fluctuations due to bursty star formation (of the sort seen in our simulations here) produce cusp-core transformations \citep[e.g.][]{2005MNRAS.356..107R,2012MNRAS.421.3464P,2015arXiv150207356P}. Such fluctiations occur on $\sim$ the scale at which stars form, which is $\sim r_{1/2}$.

Notice that core formation proceeds much more quickly in the $M_{200} = 10^8$\,M$_\odot$ simulation than in the $10^9$\,M$_\odot$ simulation. This is simply because the core is smaller and so the dynamical time within the core is smaller. The $10^8$\,M$_\odot$ simulation forms almost no stars after $\sim 4$\,Gyrs -- the SNe feedback is sufficient to shut down star formation without appealing to any external environmental agent. Yet core formation is already complete by this time. By contrast, the $5 \times 10^8$\,M$_\odot$ and $10^9$\,M$_\odot$ simulations continue to form stars for a Hubble time, and their cores continue to grow and flatten over this time. Interestingly, our $10^7$\,M$_\odot$ simulation did not form enough stars for core formation to complete; the final dark matter density profile after 14\,Gyrs of evolution is very close to the initial NFW density profile. However, we lack the resolution and physics to trust this result (only two star particles form). For this reason, we do not show the results for this simulation here. We will return to the question of whether dwarf galaxies are expected to be cusped or cored -- or even exist at all -- in $10^7$\,M$_\odot$ halos, in future work.

In the following sub-sections, we discuss the sensitivity of our results to our choice of numerical `sub-grid' parameters and initial conditions; and we present a convenient fitting function that captures the size and time evolution of this dark matter core as a function of the star formation time and the {\it projected} stellar half mass radius $R_{1/2}$. (We use the projected half mass radius rather than $r_{1/2}$ because $R_{1/2}$ is more useful observationally.)

\subsubsection{Sensitivity to our numerical `sub-grid' parameters}\label{sec:sensitivity} 

We have asserted so far that at the adopted minimum cell size of $\Delta x=4$ pc we should become largely insensitive to our choice of `sub-grid' numerical parameters. In this section, we explicitly test this by running a suite of simulations at fixed halo mass ($M_{\rm 200}=10^8$\,M$_\odot$) while varying a number of key parameters in relation to the fiducial simulation (see also Table\,\ref{tab:sims}). Our suite of `sensitivity analysis' simulations differ from the fiducial run, as follows: 

\begin{itemize}
\item M8c28\_4e6\_r1e3 $\mid$ In this simulation, we increase the number of dark matter particles from $N=4\times 10^5$ to $4\times 10^6$ while raising the star formation density threshold to $\rho_* = 10^3 m_{\rm H}$\,cm$^{-3}$. In this way, we simultaneously test the effect of both resolution and sensitivity to $\rho_*$.  
\item M8c28\_4e5\_e001 $\mid$ In this simulation, we decrease the local star formation efficiency per free-fall time (\S\,\ref{sec:SF}) from $\epsilon_{\rm ff}=10\%$ to $1\%$ \citep[e.g.][]{krumholztan07};
\item M8c28\_4e5\_pST $\mid$ In this simulation, we explicitly inject the momentum ($p_{\rm ST}$) generated during the Sedov-Taylor phase for every SN explosion (see discussion in \S\,\ref{sec:FB}), in addition to the initial thermal energy. This amounts to an over-injection of momentum from SN feedback, but is nonetheless useful to test our sensitivity to feedback parameters.
\end{itemize}

\begin{figure}
\begin{center}
\includegraphics[width=0.35\textwidth]{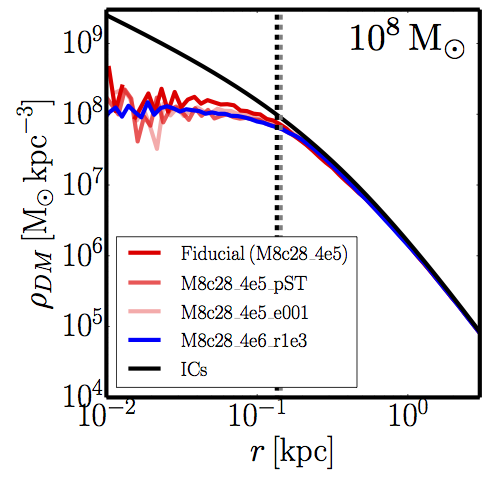}
\caption{Sensitivity of dark matter cusp-core transformations to the choice of simulation parameters. Here we re-run our fiducial simulation at $10^8$\,M$_\odot$ (red; M8c28\_4e5), explicitly injecting the momentum ($p_{\rm ST}$) generated during the Sedov-Taylor phase for every SN explosion (light red; M8c28\_4e5\_pST); decreasing the local star formation efficiency per free-fall time from $\epsilon_{\rm ff}=10\%$ to $1\%$ (pink; M8c28\_4e5\_e001); and increasing the number of dark matter particles by a factor 10 while raising the star formation density threshold to $\rho_* = 10^3\,m_{\rm H}$\,cm$^{-3}$ (blue; M8c28\_4e6\_r1e3). The initial NFW density profile is marked in black. The vertical grey and black dashed lines mark the 3D stellar half mass radius ($r_{1/2}$) and 2D projected stellar half mass radius ($R_{1/2}$), respectively. Notice that none of these changes produces a significant change in the central dark matter density profile; in all cases there is a dark matter core of size $\sim r_{1/2}$.}
\label{fig:sensitivity} 
\end{center}
\end{figure}

The resulting, almost identical, dark matter density profiles, after 14\,Gyr of evolution and for each of the above parameter variations, are shown in Figure \ref{fig:sensitivity}. This illustrates that order of magnitude changes in simulation parameters have no significant effect on our conclusions \citep[see also e.g.][]{2008PASJ...60..667S,2011MNRAS.417..950H}.

In addition to the above tests, we have have rerun simulations where we record the gas densities during SNe events in order to quantify the fraction of resolved SNe events, see discussion in \S\,\ref{sec:FB}. In the case of the M5e8c25\_2e6 simulation, we satisfy the $r_{\rm cool}\geq3\Delta x$ resolution criterion for $97\%$ of all SNe events. The reason for this is that pre-SN feedback, here radiation pressure and stellar winds, as well as gas turbulence, clears star forming regions of dense gas before most SNe events occur (up to $\sim 40$ Myr after star formation). Even though the adopted star formation threshold in our fiducial suite of simulations is $300\,{m_{\rm H}\,{\rm cm}^{-3}}$, only 26 out of almost $\sim 55,000$ explosions occur at $m\ge 300\,m_{\rm H}\,{{\rm cm}^{-3}}$, with the absolute majority ($\sim 99\%$) occurring at densities $n<1\,m_{\rm H}\,{\rm cm}^{-3}$.

\subsubsection{Sensitivity to our choice of initial conditions}\label{sec:sensitivityICs} 

\begin{figure*}
\begin{center}
\includegraphics[width=0.95\textwidth]{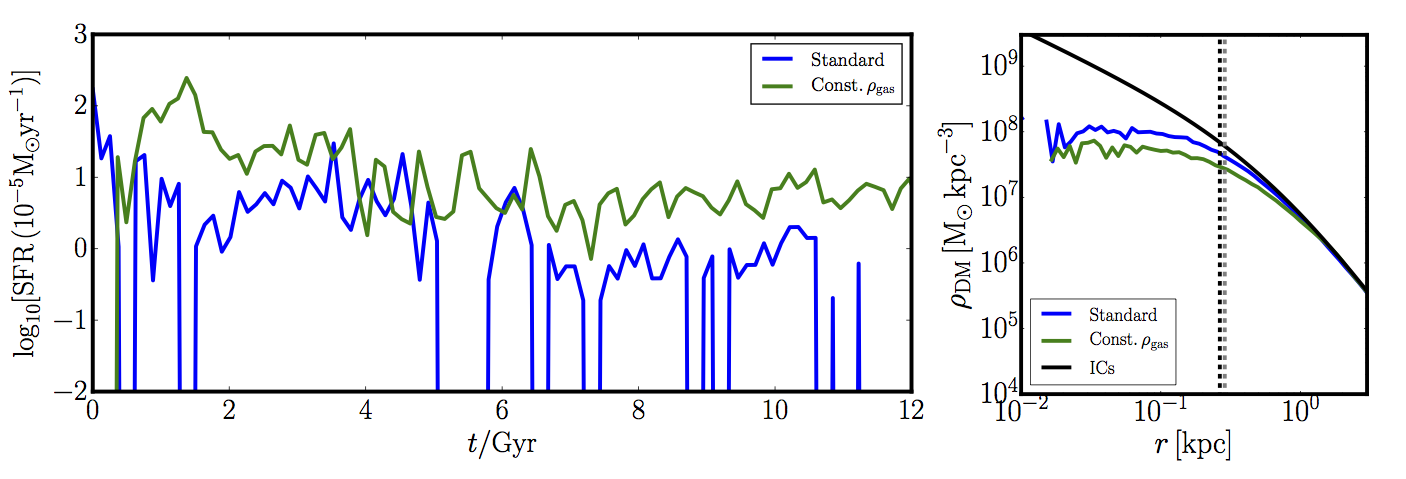}
\caption{Sensitivity of dark matter cusp-core transformations to the choice of initial conditions. Here we rerun our fiducial $M_{200} = 5\times 10^8$\,M$_\odot$ simulation (blue) using a constant initial gas density (M5e8c25\_2e6\_rhocon; green). The left panel compares the star formation rate as function of time; the right panel shows the resulting change in the dark matter density profile at the end of the simulation.}
\label{fig:sensitivityICs} 
\end{center}
\end{figure*}

In this section, we test the robustness of our results to our choice of initial conditions. Recall that our simulations are somewhat idealised in that we start with a fully formed NFW dark matter halo, filled with the universal baryon fraction of gas. For our fiducial runs, the gas was set up as an NFW profile in hydrostatic equilibrium (see \S\ref{sec:ics}). Here, we explore the effect of this assumption in an additional simulation -- M5e8c25\_2e6\_rhocon. This is identical to our $M_{200} = 5\times 10^8$\,$M_\odot$ fiducial run but initialised with constant gas density out to $r_{200}$ (see Table \ref{tab:sims}). The results are shown in Figure \ref{fig:sensitivityICs}. The left panel compares the star formation rate as function of time; the right panel shows the resulting change in the dark matter density profile at the end of the simulation.

Firstly, notice that M5e8c25\_2e6\_rhocon forms no stars for the first $\sim 0.4$\,Gyr after which it has a substantial starburst at $\sim 1$\,Gyr. This occurs because the constant density slab must first cool to form stars, unlike our fiducial run M5e8c25\_2e6 that is initialised with a central density cusp. Once the constant density slab has cooled, however, it provides a much higher accretion rate of gas than our fiducial run since it is initially much cooler and denser at large radii. This drives a higher star formation rate for the whole simulation, with a correspondingly larger final stellar mass of $M_* = 2.3 \times 10^6$\,M$_\odot$ as compared to our fiducial run that forms just $M_* = 0.6 \times 10^6$\,M$_\odot$. This demonstrates that our results are sensitive to the cosmic gas accretion history. We will study this further in future work where we will model isolated dwarfs in their cosmological context. Our main result, however, that dark matter cores of size $\sim r_{1/2}$ form, is robust. In Figure \ref{fig:sensitivityICs}, we see that both our fiducial run and M5e8c25\_2e6\_rhocon form cored dark matter profiles at the end of the simulation, of size $\sim r_{1/2}$. For M5e8c25\_2e6\_rhocon, the central dark matter density is lower by a factor $\sim 2$ as compared to our fiducial run, reflecting the larger energy input from its larger stellar mass (see also \S\ref{sec:energy}).

Finally, we stress that M5e8c25\_2e6\_rhocon was deliberately chosen to be extreme in order to test the maximal sensitivity to our initial conditions. In $\Lambda$CDM, dark matter halos assemble primarily through successive mergers \citep{1978MNRAS.183..341W}. It is highly unlikely that following such an assembly, the gas would be arranged as a cold constant density slab. At leading order, the gas should trace the underlying dark matter as in our fiducial setup \citep[e.g.][]{2015ApJ...808...40W}.

\subsubsection{A convenient \coreNFW\ fitting function}\label{sec:coreNFW}

In this section, we derive a convenient fitting function to capture the time evolution of our dark matter halos: the \coreNFW\ profile. We find that the spherically averaged dark matter profile evolution of our dwarfs is well-characterised by a modified NFW functional form: 

\begin{equation}
M_{\rm cNFW}(<r) = M_{\rm NFW}(<r) f^n
\end{equation}
where $M_{\rm NFW}(<r)$ is as in equation \ref{eqn:MNFW}; and the function $f^n$ generates a shallower profile below a core radius $r_c$: 

\begin{equation} 
f^n = \left[\tanh\left(\frac{r}{r_c}\right)\right]^n
\end{equation}
where the parameter $0 < n \le 1$ controls how shallow the core becomes, where $n=0$ corresponds to no core and $n=1$ to complete core formation. We tie the parameter $n$ to the total star formation time $t_{\rm SF}$:
\begin{equation} 
n = \tanh(q) \,\,\,\, ; \,\,\,\, q = \kappa \frac{t_{\rm SF}}{t_{\rm dyn}}
\label{eqn:kappaqn}
\end{equation} 
where $t_{\rm dyn}$ is the circular orbit time at the NFW profile scale radius $r_s$ (see equation \ref{eqn:rhoNFW}): 

\begin{equation} 
t_{\rm dyn} = 2\pi \sqrt{\frac{r_s^3}{G M_{\rm NFW}(r_s)}}
\end{equation} 
Note that the `star formation time' $t_{\rm SF}$ here simply means the length of time that the simulation has run for (since stars form at a near continuous rate). 

The dark matter core size is set by the projected stellar half mass radius of the stars: 

\begin{equation} 
r_c = \eta R_{1/2}
\label{eqn:etarc}
\end{equation}
which leaves us with just two tuning parameters $\eta$ and $\kappa$. We find $\kappa = 0.04$ and $\eta = 1.75$ give a good characterisation of our simulation results over the full mass range (see the thin dashed lines in Figure \ref{fig:cuspcoretransforms}). Notice that in the limit $t_{\rm SF} \rightarrow 0$, $M_{\rm cNFW} \rightarrow M_{\rm NFW}$ and we recover the usual NFW profile. Similarly for $r \gg r_c$, we return to the NFW form. This is advantageous as it means that our fitting function conserves NFW profile mass; while the virial mass $M_{200}$ and concentration parameter $c$ (see equation \ref{eqn:rhoNFW}) take on their usual meanings.

The density profile follows from the radial derivative of the mass profile as: 

\begin{equation} 
\rho_{\rm cNFW}(r) = f^n \rho_{\rm NFW} + \frac{n f^{n-1} (1-f^2)}{4\pi r^2 r_c} M_{\rm NFW}
\end{equation} 
Our functional form introduces two new parameters as compared to the NFW profile -- $t_{\rm SF}$ and $R_{1/2}$ -- that control the core flattening and size, respectively. Both are {\it observable}, however, allowing our derived density profile to be readily compared to real data. We will consider such comparisons in forthcoming papers.
 
\subsection{Scatter in the \coreNFW\ tuning parameters} 

\begin{figure}
\begin{center}
\includegraphics[width=0.35\textwidth]{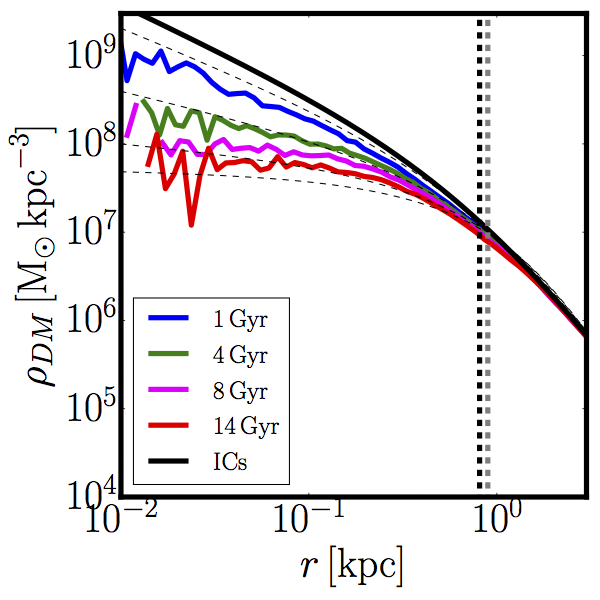}
\caption{Sensitivity of dark matter cusp-core transformations to the halo spin parameter $\lambda'$. Here we shows results for the dark matter density profile evolution in simulation M9c22\_4e6\_lam007 (see Table \ref{tab:sims}; the lines are as in Figure \ref{fig:cuspcoretransforms}). This is identical to our fiducial $M_{200} = 10^9$\,M$_\odot$ simulation but with twice the initial dark matter halo spin parameter ($\lambda' = 0.07$), resulting in a larger projected half stellar mass radius. Notice that the \coreNFW\ profile shown by the thin dashed lines qualitatively captures the evolution of the dark matter density profile, but the match is poorer than for the fiducial simulation in Figure \ref{fig:cuspcoretransforms}. This suggests that the \coreNFW\ tuning parameters $\eta$ and $\kappa$ have some weak dependence on halo spin $\lambda'$ and concentration $c$; we defer further analysis of this to future work.}
\label{fig:dmproflam007} 
\end{center}
\end{figure}

Our \coreNFW\ profile introduces two new parameters that are observable: the total star formation time $t_{\rm SF}$ and the projected stellar half mass radius $R_{1/2}$; and two new parameters that are tuned to fit the simulations: $\eta$ that controls the size of the dark matter core via equation \ref{eqn:etarc} and $\kappa$ that controls how rapidly the core forms via equation \ref{eqn:kappaqn}. One might reasonably ask whether these latter two show some scatter with halo concentration $c$, spin parameter $\lambda'$ and/or assembly history. Our small sample of simulations is not sufficient to fully answer this question, but we can obtain some hint using our high spin simulation: M9c22\_4e6\_lam007 (Table \ref{tab:sims}). This is identical to our fiducial $M_{200} = 10^9$\,M$_\odot$ simulation but with double the cosmic mean spin parameter. The observable properties of this simulation are qualitatively very similar to our fiducial simulation. However, perhaps unsurprisingly, it has a larger projected stellar half mass radius ($R_{1/2} = 0.81$\,kpc as compared to $R_{1/2} = 0.6$\,kpc for the fiducial simulation). This allows us to test how well the \coreNFW\ form works for rather extreme changes in the spin parameter. (Note that $\lambda' = 0.07$, as assumed in M9c22\_4e6\_lam007, is relatively rare; $\sim 8$\% of halos will have $\lambda' > 0.07$; e.g. \citealt{2001ApJ...555..240B}.) The results are shown in Figure \ref{fig:dmproflam007}. Notice that the \coreNFW\ profile shown by the thin dashed lines qualitatively captures the evolution of the dark matter density profile, but the match is poorer than for the fiducial simulation in Figure \ref{fig:cuspcoretransforms}. This suggests that the \coreNFW\ tuning parameters $\eta$ and $\kappa$ have some weak dependence on halo spin $\lambda'$ and concentration $c$; we defer further analysis of this to future work.

\section{Discussion}\label{sec:discussion}

\subsection{The energetics of cusp-core transformations}\label{sec:energy}

Several papers in the literature have argued that cusp-core transformations should become energetically unfavourable below some critical stellar or dark matter mass \citep[e.g.][and see discussion in \S\ref{sec:introduction}]{2012ApJ...759L..42P,2013MNRAS.433.3539G}. Yet here, we argue that cores always form if star formation proceeds for long enough. In this section, we explain why we arrive at a different result.

Firstly, let us check that our simulations make sense from simple energetics arguments. The difference in gravitational binding energy of our \coreNFW\ profile with respect to the NFW profile can be calculated as: 

\begin{equation} 
\Delta W = -\frac{1}{2}\int_0^\infty \frac{G(M_{\rm NFW}^2 - M_{\rm cNFW}^2)}{r^2} dr
\end{equation}
which is the amount of energy required to transform our initially cuspy profiles to cored profiles. We compare this with the available energy from SNe in Table \ref{tab:energetics}. As can be seen, in all cases $< 1$\% of the available SNe energy is required to unbind the cusp. Core formation does take time, however. If star formation is truncated then there will not have been enough integrated SNe energy to unbind the cusp, the extreme example being the case where we truncate star formation after a single star has formed. In this sense, it is absolutely correct that, for a given halo mass, there is a critical {\it stellar mass} at which core formation becomes energetically impossible. Our simulations suggest, however, that there is no critical {\it halo mass} at which this is the case -- at least not down to $10^8$\,M$_\odot$. It is possible that core formation ceases at $\sim 10^7$\,M$_\odot$ because so few stars form, but the simulations we present here do not resolve this mass scale well enough to be able to draw strong conclusions. Below this mass scale, however, we expect galaxy formation to be extremely challenging. If $\Lambda$CDM is correct, we would expect dark matter halos below $\sim 10^7$\,M$_\odot$ to be almost devoid of stars and, therefore, to retain their pristine dark cusp.

Let us now connect explicitly to the calculation in \citet{2012ApJ...759L..42P}. Their equation 6 gives the supernovae energy used to transform cusps to cores as: 

\begin{equation} 
\frac{\Delta E}{E_{\rm SN}} = \frac{M_*}{\langle m_* \rangle} \xi(m_* > 8\,{\rm M}_\odot) \epsilon_{\rm DM}
\label{eqn:pencalc}
\end{equation}
where $\langle m_* \rangle$ and $\xi$ are the mean stellar mass and the fraction of mass in stars that go supernova, respectively; and $\epsilon_{\rm DM}$ is the efficiency of coupling of the SNe energy to the dark matter (i.e. the same quantity that appears in Table \ref{tab:energetics}). Both $\langle m_* \rangle$ and $\xi$ depend on the assumed IMF. For the \cite{chabrier03} IMF we assume here (see \S\ref{sec:FB}), these are $\langle m_* \rangle = 0.83$ (averaged over the range $0.1 < m_*/{\rm M}_\odot < 100$)  and $\xi = 0.00978$. With these numbers, and using $M_{*,{\rm birth}}$ (i.e. $M_*$ corrected for mass loss due to stellar evolution), we find: $\Delta E / E_{\rm SN} = [3.8, 34, 160]$ for $M_{200} = [10^8, 5\times 10^8, 10^9]\,{\rm M}_\odot$, respectively. This gives excellent agreement with the $\Delta W / E_{\rm SN}$ reported in Table \ref{tab:energetics}, as it should.

\citet{2012ApJ...759L..42P} find that core formation should become inefficient below $M_{200} \sim 10^{10}$\,$M_\odot$. We can now use equation \ref{eqn:pencalc} to understand why we find cores at much smaller mass scales than this. The first potential effect is in our different assumed IMF. If we switch to the \citet{2002Sci...295...82K} IMF assumed by \citet{2012ApJ...759L..42P}, we have $\langle m_* \rangle = 0.4$ and $\xi = 0.0037$. Since it is the ratio of these quantities that matters, this makes only a small difference, lowering the available SNe energy by $\sim 20$\%. The second effect is the assumed core size $r_c$ which affects the amount of SNe energy required for core formation ($\Delta W$). \citet{2012ApJ...759L..42P} assume either $r_c = 1$\,kpc or $r_c = 0.1 r_s$. For our $10^9$\,M$_\odot$ halo, $r_c = 0.83$\,kpc, while $0.1 r_s = 0.1$\,kpc so our dark matter cores sit comfortably within the range assumed by \citet{2012ApJ...759L..42P}. Then there is the energy coupling efficiency $\epsilon_{\rm DM}$. \citet{2012ApJ...759L..42P} assume $\epsilon_{\rm DM} = 0.4$ which is substantially larger than we find here. However, this should act to make core formation much easier rather than harder. All of this leaves just one variable left: the stellar mass for a given halo mass. \citet{2012ApJ...759L..42P} assume a stellar mass to halo mass relation taken from the low mass extrapolation of the \citet{2010ApJ...710..903M} abundance matching relation. While this relation is extrapolated, it does match very well a recent abundance matching measurement for the Local Group by \citet{2014ApJ...784L..14B} that reaches down to $M_* \sim 10^6$\,M$_\odot$. The \citet{2010ApJ...710..903M} relation assigns a stellar mass of just $M_* \sim 5 \times 10^4$\,M$_\odot$ to a $M_{200} \sim 10^{9}$\,M$_\odot$ galaxy -- substantially less than the $M_* \sim 3.5 \times 10^6$\,M$_\odot$ that our simulated dwarf forms. It is clear, then, that our simulated dwarfs do not match the low mass extrapolation of the \citet{2010ApJ...710..903M} relation. This is the key reason why we are able to form cores `all the way down', unlike \citet{2012ApJ...759L..42P}. Is this a problem, however? We have shown already that our dwarfs are remarkably realistic, giving a good match to all known data for real isolated dwarfs (\S\ref{sec:observations}). Furthermore, a recent independent `tidal' mass estimate for the Carina dwarf spheroidal -- that has a stellar mass of $M_* = 4.8 \pm 0.5 \times 10^5$\,M$_\odot$ -- puts its mass before infall onto the Milky Way at just $M_{200} = 3.6^{+3.8}_{-2.3}\times 10^8$\,M$_\odot$ \citep{2015NatCo...6E7599U}. Thus, Carina's stellar mass and halo mass are in excellent agreement with the simulations that we present here. And -- just like our simulations -- Carina is in strong conflict with the low mass extrapolation of the \citet{2010ApJ...710..903M} relation. As pointed out by \citet{2015NatCo...6E7599U}, this likely means that abundance matching fails below $\sim 10^{10}$\,M$_\odot$, at least inside the Local Group. After all, we know that most of the Local Group dSphs have had their star formation truncated due to ram pressure and tides \citep[e.g.][]{2012AJ....144....4M}. This will act to lower $M_*$ for a given pre-infall $M_{200}$, destroying the monotonic relation between $M_*$ and $M_{200}$ that is an implicit assumption of the abundance matching machinery. We will present a much more detailed comparison of our simulations with abundance matching constraints in a separate companion paper.

\begin{table}
\begin{center}
\begin{tabular}{l c c c}
$M_{200}/{\rm M}_\odot$ & $10^8$ & $5 \times 10^8$ &	$10^9$ \\
\hline
$N_*$ & 86 & 2721 & 15,390 \\
$M_*/{\rm M}_\odot$ & $2 \times 10^4$ & $6.2 \times 10^5$ & $3.6 \times 10^6$ \\
$M_{*,{\rm birth}}/{\rm M}_\odot$ & $4 \times 10^4$ & $12.6 \times 10^5$ & $7.13 \times 10^6$ \\
$N_{\rm SN}$ & 465 & 14,761 & 83,483 \\
$\Delta W / E_{\rm SN}$ & 3.8 & 33.4 & 161.9 \\
$\epsilon_{\rm DM}$ & 0.8\% & 0.23\% & 0.19\%
\end{tabular}
\end{center}
\caption{Supernova energy coupling efficiency required to produce the cusp-core transformations in our fiducial simulations. The rows give the dark matter halo mass of the simulated dwarf ($M_{200}$); the number of star particles formed $N_*$; the total stellar mass $M_*$; the total stellar mass at birth $M_{*,{\rm birth}}$ (i.e. before mass loss due to stellar evolution); the total number of SNe $N_{\rm SN}$; the gravitational binding energy required to transform the cusp to a core $\Delta W$, in units of the SN energy ($E_{\rm SN} = 10^{51}$\,ergs); and the coupling efficiency $\epsilon = \Delta W / (E_{\rm SN} N_{\rm SN})$. Notice that for all simulations, the coupling efficiency is $< 1$\%.}
\label{tab:energetics}
\end{table}

\subsection{Implications for the cusp-core problem: where to find pristine cusps}

Our results in Figure \ref{fig:cuspcoretransforms} suggest that -- if $\Lambda$CDM is correct -- `pristine' dark matter cusps will be found either in galaxies that have truncated star formation\footnote{Note that intermittent or `choked' star formation would also significantly damp dark matter core formation, as might occur in some reionisation models \citep[e.g.][]{2009MNRAS.392L..45R}, or if dwarfs are not fully ram pressure stripped in a single pericentre passage.}, or at radii $r > r_{1/2}$. This amounts to a strong prediction of the theory that can be distinguished from modifications to dark matter, since modified dark matter models will act in the same way on all halos irrespective of their star formation history \citep[e.g.][]{2013MNRAS.430...81R}. \citet{2015arXiv150202036O} already noted that it takes time for cores to form. Similarly to our results here, they find that galaxies with truncated star formation should be more cuspy. Here we add two additional key points: (i) cores should always be of order the stellar half mass radius and thus `pristine' cusps may also be found at $r > r_{1/2}$ (provided $r_{1/2} < r_s$); and (ii) we provide a fitting function -- the \coreNFW\ profile -- that predicts the cusp-core evolution as a function of star formation time. This allows us to compare the model with data even for galaxies that are transitioning between being cusped and cored. We will consider such comparisons between model and data in a series of forthcoming papers. 

\subsection{Implications for the missing satellites problem: tidal scouring of dwarfs}

Notice that after cusp-core transformation, the central dark matter density of all of our dwarfs is almost the same with $\rho(0) \sim 10^8$\,M$_\odot$\,kpc$^{-3}$ (see Figure \ref{fig:cuspcoretransforms}). This is important since such low density dwarfs will be very susceptible to tidal forces when falling into a larger host galaxy \citep[e.g.][]{2006MNRAS.366..429R}. Such physics will be necessarily missed by simulations that model only the dark matter fluid since they do not include the physics required to model cusp-core transformations \citep{2006MNRAS.tmp..153R,2010MNRAS.406.1290P,2013ApJ...765...22B}.

We will consider the effect of tides in detail on these dwarfs in a forthcoming publication. Here, we use the tidal radius formula from \citet{2006MNRAS.366..429R} to estimate at what mass and orbital pericentre such cored dwarfs will be completely destroyed when falling into the Milky Way:

\begin{eqnarray}
0 & = & \frac{GM_g(x)}{x^2} - \frac{GM_g(x-r_t)}{(x-r_t)^2} + \frac{GM_s(r_t)}{r_t^2} - \\
& & \Omega^2 r_t - 2 \alpha \Omega \Omega_s r_t
\label{eqn:finaleqrea}
\end{eqnarray}
\noindent
where $M_s(r)$ and $M_g(x)$ are the satellite and host galaxy mass distributions, respectively; $r_t$ is the satellite tidal radius; $x$ is the distance from the host to the satellite; $\Omega = J/x^2$ is the angular velocity of the satellite of specific angular momentum $J$ about the host; $\Omega_s^2 = GM_s(r_t)/r_t^3$ is the angular velocity of stars within the satellite; and the parameter $\alpha$ corresponds to the orientation of orbits within the satellite with respect to its orbit about the host: 

\begin{equation}
\alpha = \left\{\begin{array}{ll}
      1 & \mbox{prograde} \\
      0 & \mbox{radial} \\
     -1 & \mbox{retrograde}
     \end{array}
     \right.
\label{eqn:alpha}
\end{equation}
here we assume $\alpha = 1$, while the specific angular momentum of the satellite follows from the orbital peri- and apocentre: 

\begin{equation} 
J^2 = 2\frac{(\Phi_g(x_p) - \Phi_g(x_a))x_a^2x_p^2}{x_p^2-x_a^2}
\end{equation}
where $\Phi_g$ is the gravitational potential of the host galaxy.

Similarly to \citet{2009ApJ...706L.150A}, we crudely approximate the `Milky Way' halo as a spherical Hernquist profile with mass and scale length $M_h = 10^{12}$\,M$_\odot$; $r_h = 20$\,kpc. For the satellite we assume a \coreNFW\ profile. We define the satellite as being `destroyed' if the tidal radius is smaller than the dark matter core radius: $r_t < r_c$. The core radius is set by the projected stellar half mass radius $r_c = 1.75 R_{1/2}$ (c.f. \S\ref{sec:cuspcore}). We find that the relationship between $R_{1/2}$ and $M_{200}$ for our simulations agrees within a factor of two with that found in \citet{2013ApJ...764L..31K,2015ApJS..219...15S,2015arXiv150900853A}, namely: 

\begin{equation} 
R_{1/2} \sim r_{1/2} \sim 0.015 r_{200}
\end{equation}
where $r_{200}$ is related to the virial mass $M_{200}$ via equation \ref{eqn:r200}. Thus, the core radius $r_c \propto M_{200}^{1/3}$ grows weakly with increasing mass. 

With the above approximations, we find that satellites are destroyed almost independently of mass and orbital apocentre up to $\sim 10^{10}$\,M$_\odot$; all satellites with $r_p \simlt 30$\,kpc and $M_{200} \simlt 10^{10}$\,M$_\odot$ will be tidally shredded rapidly after infall onto the Milky Way. The only way to avoid this is to truncate star formation and maintain a steeper central cusp. Indeed, the only reason why many low mass satellites are found around the Milky Way at all may be because their star formation was truncated rapidly by ram pressure stripping. Recent models suggest that ram pressure can remove almost all of the ISM of a dwarf in just one pericentric passage \citep{2013MNRAS.433.2749G}. This may be vital for the survival of low mass dwarfs that orbit close to our Galaxy. Indeed, \citet{2010MNRAS.406.1290P} noted already that if the ultra-faint dwarfs were cored they would not survive many orbits at their current locations in the Galaxy\footnote{\citet{2014MNRAS.438.1466A} point out that ram pressure stripping of gas can also lower the central stellar and dark matter density of satellites (see also a similar discussion in \citealt{2005MNRAS.356..107R}). This effect must be relatively small, however, if dSphs are to survive within the harsh tidal field of the Milky Way and/or Andromeda.}.

Such tidal scouring of dwarfs will dramatically reshape the subhalo mass function inside the Milky Way (and similarly for other large spiral or group environments). Surviving subhalos will be those either on benign orbits that keep them far from the Milky Way, or those that fell in early and had their star formation truncated (and thus managed to maintain a steep central density cusp). In this context, it is perhaps telling that the two low mass dwarfs in the Milky Way with extended star formation (Carina and Fornax) have large orbital pericentres \citep{2010arXiv1001.1731L}. While the errors are large and proper motion measures fraught with difficulty, the latest data point to Fornax being on a cosmologically bizarre near-circular orbit; while Carina has a median $r_p \sim 45$\,kpc. Typical orbits in $\Lambda$CDM should have apo-to-peri ratios closer to 1:5 for the surviving satellites and 1:6 for the accreted ones \citep[e.g.][]{2007arXiv0708.1949K,2008MNRAS.389.1041R,2010arXiv1001.1731L}. The fact that the Milky Way classical dSphs have such seemingly circular orbits can be explained if all the dwarfs on more eccentric orbits were destroyed by tides. This then naturally alleviates the missing satellites problem, but in a manner very different to most other explanations in the literature. Typical solutions involve various prescriptions for painting stars inefficiently on dark matter subhalos \citep[e.g.][]{2009ApJ...696.2179K}; in the picture we present here most subhalos will simply cease to exist. Such a scenario was explored recently by \citet{2013ApJ...765...22B} where they show that it can indeed solve the missing satellites problem. 

There are three final implications of the above worth highlighting. Firstly, since subhalos are actually destroyed this will affect methods like lensing \citep[e.g.][]{2002ApJ...567L...5M,2009arXiv0909.5133B} or `satellite stream bumps' \citep[e.g.][]{2002ApJ...570..656J} that gravitationally probe the existence of purely dark satellites. Both should find a subhalo mass function that is depleted (at least over the mass range $10^8 \simlt M_{200}/{\rm M}_\odot \simlt 10^{10}$) as compared to predictions from pure dark matter simulations. Secondly, while surviving dwarfs may be on relatively benign orbits, their dark matter cores can still play a role in facilitating their transformation from dwarf irregulars to dwarf spheroidals \citep{2001ApJ...547L.123M,2012ApJ...751L..15L,2013ApJ...764L..29K}. In particular, \citet{2013ApJ...764L..29K} show that cored dwarfs on `benign' orbits can still be efficiently transformed by tides. Thirdly, since both low and high mass satellites are efficiently destroyed this will also have an impact on the `too big to fail' problem. We discuss this further, next.

\begin{figure*}
\begin{center}
\includegraphics[width=0.49\textwidth]{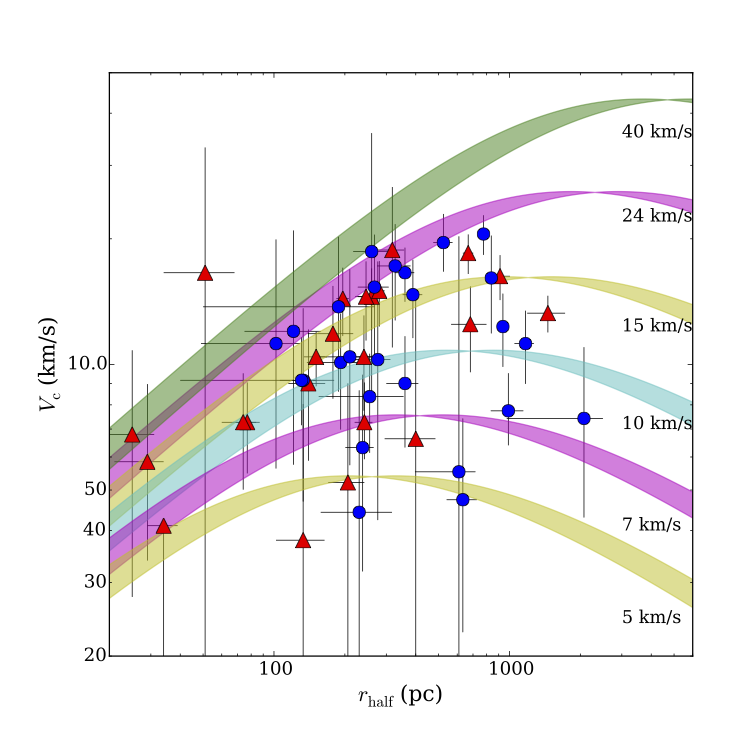}
\includegraphics[width=0.49\textwidth]{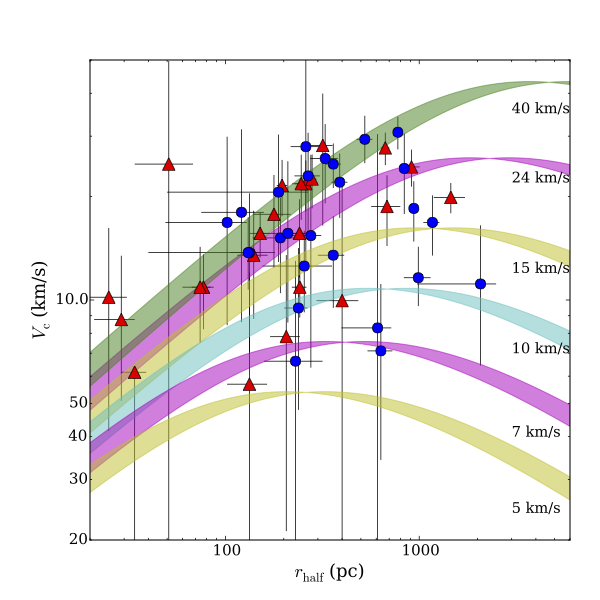}
\caption{Measured circular velocities within the half light radius of the Local Group dSphs, derived as in \citet{2014ApJ...783....7C}. In the left panel, we use the measured velocity dispersions; in the right panel, we raise the measured dispersions by a factor 1.5 to compensate for (incomplete) cusp-core transformations. (Recall that cusp-core transformations can shift the central velocity dispersion of dwarfs by up to a factor of two; see Figures \ref{fig:sim_data_5e8}(f) and \ref{fig:sim_data_1e9}(f).) The coloured bands in both panels mark dark matter halo circular speed curves taken from \citet{2011MNRAS.415L..40B}; the peak circular velocity $v_{\rm max}$ is marked in each case. According to \citet{2011MNRAS.415L..40B}, there should be roughly 10 dwarf galaxies around the MW(/M31) that live in such massive halos. With the factor of 1.5 uniformly applied to the measured velocity dispersions, we end up with 13 MW dSphs (red) and 14 M31 dSphs (blue) that live in $v_{\rm max} \sim 40$\,km/s halos. This is sufficient to completely solve the `too big to fail' problem.}
\label{fig:vcirc_sigmaboost} 
\end{center}
\end{figure*}

\subsection{Implications for `too big to fail'}

Once cusp-core transformations are taken into account, many massive subhalos will be tidally shredded on infall to the Milky Way, as discussed above. However, subhalos that fall in early enough will have their star formation truncated before core formation is complete, allowing them to survive. Such early infalling halos are known to be preferentially those that form early at high redshift \citep[e.g.][]{2005MNRAS.364..367D}. Indeed, it has already been noted that the spatial and orbit distribution of the surviving dwarfs is more consistent with these early biased peaks than that of the full subhalo distribution \citep{2005MNRAS.364..367D,2005astro.ph.10370M,2010arXiv1001.1731L}. This is usually attributed to dwarfs forming more easily before reionisation. However, several authors have argued that once satellites obtain some cold gas, they can efficiently self-shield against reionisation and continue to form stars \citep[e.g.][]{2000ApJ...542..535G,2004ApJ...600....1S}. It is perhaps telling that no unambiguous reionisation feature in the star formation history of surviving dwarfs has been reported to date, despite many studies looking for one \citep[e.g.][]{2007ApJ...659L..17C,2011ApJ...730...14H,2014ApJ...786...44S,2014ApJ...789..148W}. This suggests another explanation for why biased peaks are favoured as survivors. Here, we suggest that it is their early infall that matters. This causes their star formation to be truncated early by ram pressure, leading to the survival of their dark matter cusps. This then allows them to survive in the tidal field of the Milky Way.

This still leaves us with a puzzle, however. If the classical dwarfs are early infallers that had their star formation truncated then we seem to return once more to the `too-big-to-fail' problem. \citet{2011MNRAS.415L..40B} show that even early infalling subhalos are too massive to be consistent with the Milky Way classical dSphs. However, cusp-core transformations once again have an important role to play. Firstly, dwarfs on benign orbits like Carina and Fornax can form stars for a Hubble time and undergo complete core formation. Secondly, even dwarfs on more extreme orbits could undergo partial core formation. As shown in \S\ref{sec:cuspcore}, complete core formation lowers the central stellar velocity dispersion by a factor $\sim 2$ inside the stellar half mass radius. In Figure \ref{fig:vcirc_sigmaboost}, we explore the effect of this on the `too big to fail' problem. We plot the circular velocities within the half light radius of the Local Group dSphs, derived as in \citet{2014ApJ...783....7C}. In the left panel, we use the measured velocity dispersions; in the right panel, we artificially raise the measured dispersions by a factor 1.5 to crudely `undo' the effect of (incomplete) cusp-core transformations. The coloured bands in both panels mark dark matter halo circular speed curves taken from \citet{2011MNRAS.415L..40B}; the peak circular velocity $v_{\rm max}$ is marked in each case. 

As can be seen in Figure \ref{fig:vcirc_sigmaboost}, compensating for cusp-core transformations (right panel) leads to many of the Milky Way (red) and Andromeda (blue) dwarfs becoming consistent with inhabiting $v_{\rm max} \sim 40$\,km/s halos. According to \citet{2011MNRAS.415L..40B}, there should be roughly 10 dwarf galaxies around the MW(/M31) that live in such massive halos. With the factor of 1.5 uniformly applied to the measured velocity dispersions, we end up with 13 MW dSphs (red) and 14 M31 dSphs (blue) that live in $v_{\rm max} \sim 40$\,km/s halos. This is sufficient to completely solve the `too big to fail' problem. Whether this works in detail will require more sophisticated modelling of the dwarf population in its cosmological context that we defer to future work. Here, we simply point out that cored dwarfs have colder central velocity dispersions and that this already goes a long way to solving `too big to fail', as has been emphasised already by previous authors \citep[e.g.][]{2006MNRAS.tmp..153R,2014ApJ...789L..17M}.

\section{Conclusions}\label{sec:conclusions} 

We have used high resolution simulations of isolated dwarf galaxies to study the physics of dark matter cusp-core transformation at the edge of galaxy formation $M_{\rm 200} = 10^7 - 10^9$\,M$_\odot$. We worked at a resolution ($\sim$4\,pc minimum cell size; $\sim$250\,M$_\odot$ per particle) at which the impact from individual supernovae explosions can be resolved, becoming insensitive to even large changes in our numerical `sub-grid' parameters. Our key results are as follows: 

\begin{itemize}

\item Dark matter cores of size comparable to the stellar half mass radius $r_{1/2}$ {\it always form} if star formation proceeds for long enough. Cores fully form in less than 4\,Gyrs for the $M_{\rm 200} =10^8$\,M$_\odot$ and $\sim 14$\,Gyrs for the $10^9$\,M$_\odot$ dwarf.

\item We provide a convenient two parameter `\coreNFW' fitting function that captures this dark matter core growth as a function of star formation time and the projected stellar half mass radius (\S\ref{sec:coreNFW}).

\item We showed that our dwarf galaxies give a remarkable match to the stellar light profile; star formation history; metallicity distribution function; and star/gas kinematics of isolated dwarf irregular galaxies (Figures \ref{fig:sim_data_5e8} and \ref{fig:sim_data_1e9}). In particular, our results suggest that the isolated dwarf galaxy Leo T has a mass $M_{200} \sim 5 \times 10^8$\,M$_\odot$, and that we are currently viewing it nearly face on (at an inclination angle of $\sim 20^\circ$). We argue that this explains the lack of an observed rotation curve for Leo T.

\item We make a strong prediction that if $\Lambda$CDM is correct, then `pristine' dark matter cusps will be found either in systems that have truncated star formation and/or at radii $r > r_{1/2}$.

\item Complete core formation lowers the projected velocity dispersion at $r_{1/2}$ by a factor $\sim 2$, which is sufficient to fully explain the `too big to fail problem' (though we stress that a full solution likely also involves unmodelled environmental effects).

\item Cored dwarfs will be much more susceptible to tides, leading to a dramatic scouring of the subhalo mass function inside galaxies and groups.
 
\item Our simulated dwarfs naturally lead to younger stars being more centrally concentrated than the older stars. This occurs because the older stars are collisionlessly heated similarly to the dark matter, being pushed out to larger radii. Such a signature is seen in the Aquarius dwarf irregular and is well-matched by our $M_{200} = 10^9$\,M$_\odot$ simulation. Similar age-radius gradients have also been reported for a much larger sample of nearly dwarf irregulars by \citet{2012AJ....143...47Z}. They interpret the signature as an `outside-in' shrinking of the star formation in these dwarfs. Here we suggest instead that it is a sign of collisionless heating caused by bursty star formation; the same heating that transforms dark matter cusps to cores. 

\end{itemize} 

\section{Acknowledgements}
JIR and OA would like to acknowledge support from STFC consolidated grant ST/M000990/1. JIR acknowledges support from the MERAC foundation. Support for MLMC was provided by NASA through Hubble Fellowship grant \#51337 awarded by the Space Telescope Science Institute, which is operated by the Association of Universities for Research in Astronomy, Inc., for NASA, under contract NAS 5-26555. This work used PyNbody for the simulation analysis ({\tt https://github.com/pynbody/pynbody}; \citealt{2013ascl.soft05002P}). All simulations were run on the Surrey Galaxy Factory. We would like to thank Jorge Pe\~narrubia and the anonymous referee for helpful feedback that improved the manuscript.

\appendix
\section{Initial conditions test}\label{sec:ictest}

\begin{figure}
\begin{center}
\includegraphics[width=0.35\textwidth]{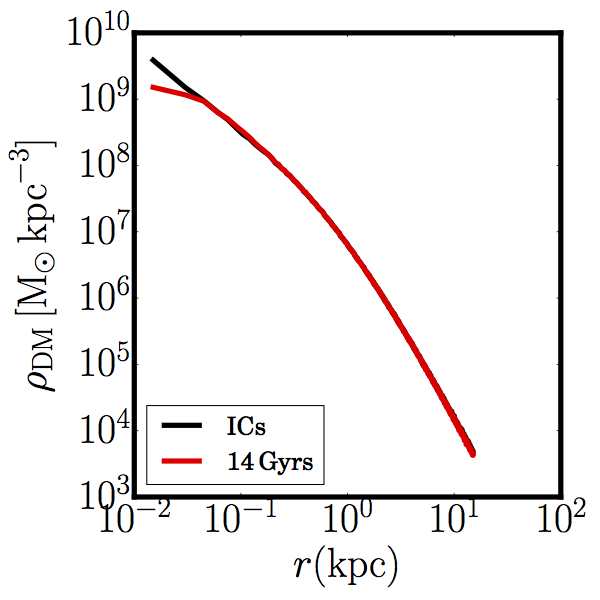}
\caption{Stability test of the dark matter initial conditions for M5e8c25\_2e6. Notice that a small core forms due to numerical relaxation; this is of size $\sim 40$\,pc which is substantially smaller than any of the dark matter cores we discuss in this paper.}
\label{fig:dm_ICtest}
\end{center}
\end{figure}

In this appendix, we test the stability of our initial conditions (see \S\ref{sec:ics} for details of the simulation set up). In Figure \ref{fig:dm_ICtest}, we show the initial (black) and final (red) spherically averaged dark matter density profile for the M5e8c25\_2e6 simulation after 14\,Gyrs of evolution, without any gas physics. This shows that our initial conditions are very stable. A small core forms due to numerical relaxation; this is of size $\sim 40$\,pc which is substantially smaller than any of the dark matter cores we discuss in this paper.

\bibliographystyle{mn2e}
\bibliography{refs,ref2}

\begin{thebibliography}{188}
\expandafter\ifx\csname natexlab\endcsname\relax\def\natexlab#1{#1}\fi

\bibitem[{{Adams} {et~al}\mbox{.}(2014){Adams}, {Simon}, {Fabricius}, {van den
  Bosch}, {Barentine}, {Bender}, {Gebhardt}, {Hill}, {Murphy}, {Swaters},
  {Thomas}, \& {van de Ven}}]{2014ApJ...789...63A}
{Adams} J.~J. {et~al.}, 2014, \apj, 789, 63

\bibitem[{{Ad{\'e}n} {et~al}\mbox{.}(2009){Ad{\'e}n}, {Wilkinson}, {Read},
  {Feltzing}, {Koch}, {Gilmore}, {Grebel}, \&
  {Lundstr{\"o}m}}]{2009ApJ...706L.150A}
{Ad{\'e}n} D., {Wilkinson} M.~I., {Read} J.~I., {Feltzing} S., {Koch} A.,
  {Gilmore} G.~F., {Grebel} E.~K., {Lundstr{\"o}m} I., 2009, \apjl, 706, L150

\bibitem[{{Agertz} \& {Kravtsov}(2015{\natexlab{a}})}]{2015ApJ...804...18A}
{Agertz} O., {Kravtsov} A.~V., 2015{\natexlab{a}}, \apj, 804, 18

\bibitem[{{Agertz} \& {Kravtsov}(2015{\natexlab{b}})}]{2015arXiv150900853A}
{Agertz} O., {Kravtsov} A.~V., 2015{\natexlab{b}}, ArXiv e-prints: 1509.00853

\bibitem[{{Agertz} {et~al}\mbox{.}(2013){Agertz}, {Kravtsov}, {Leitner}, \&
  {Gnedin}}]{2013ApJ...770...25A}
{Agertz} O., {Kravtsov} A.~V., {Leitner} S.~N., {Gnedin} N.~Y., 2013, \apj,
  770, 25

\bibitem[{{Agnello} \& {Evans}(2012)}]{2012ApJ...754L..39A}
{Agnello} A., {Evans} N.~W., 2012, \apjl, 754, L39

\bibitem[{{Amorisco} \& {Evans}(2011)}]{2011MNRAS.tmp.1606A}
{Amorisco} N.~C., {Evans} N.~W., 2011, \mnras, 1606

\bibitem[{{Amorisco} {et~al}\mbox{.}(2014){Amorisco}, {Zavala}, \& {de
  Boer}}]{2014ApJ...782L..39A}
{Amorisco} N.~C., {Zavala} J., {de Boer} T.~J.~L., 2014, \apjl, 782, L39

\bibitem[{{Arraki} {et~al}\mbox{.}(2014){Arraki}, {Klypin}, {More}, \&
  {Trujillo-Gomez}}]{2014MNRAS.438.1466A}
{Arraki} K.~S., {Klypin} A., {More} S., {Trujillo-Gomez} S., 2014, \mnras, 438,
  1466

\bibitem[{{Bacon} {et~al}\mbox{.}(2010){Bacon}, {Amara}, \&
  {Read}}]{2009arXiv0909.5133B}
{Bacon} D.~J., {Amara} A., {Read} J.~I., 2010, \mnras, 409, 389

\bibitem[{{Battaglia} {et~al}\mbox{.}(2013){Battaglia}, {Helmi}, \&
  {Breddels}}]{2013NewAR..57...52B}
{Battaglia} G., {Helmi} A., {Breddels} M., 2013, New Astronomy Reviews, 57, 52

\bibitem[{{Battaglia} {et~al}\mbox{.}(2008){Battaglia}, {Helmi}, {Tolstoy},
  {Irwin}, {Hill}, \& {Jablonka}}]{2008ApJ...681L..13B}
{Battaglia} G., {Helmi} A., {Tolstoy} E., {Irwin} M., {Hill} V., {Jablonka} P.,
  2008, \apjl, 681, L13

\bibitem[{{Begum} \& {Chengalur}(2004)}]{2004A&A...413..525B}
{Begum} A., {Chengalur} J.~N., 2004, \aap, 413, 525

\bibitem[{{Bigiel} {et~al}\mbox{.}(2008){Bigiel}, {Leroy}, {Walter}, {Brinks},
  {de Blok}, {Madore}, \& {Thornley}}]{bigiel2008}
{Bigiel} F., {Leroy} A., {Walter} F., {Brinks} E., {de Blok} W.~J.~G., {Madore}
  B., {Thornley} M.~D., 2008, \aj, 136, 2846

\bibitem[{{Binney} \& {Tremaine}(2008)}]{1987gady.book.....B}
{Binney} J., {Tremaine} S., 2008, {Galactic dynamics}. Princeton, NJ, Princeton
  University Press, 2008, 747 p.

\bibitem[{{Blondin} {et~al}\mbox{.}(1998){Blondin}, {Wright}, {Borkowski}, \&
  {Reynolds}}]{Blondin1998}
{Blondin} J.~M., {Wright} E.~B., {Borkowski} K.~J., {Reynolds} S.~P., 1998,
  \apj, 500, 342

\bibitem[{{Boley} {et~al}\mbox{.}(2009){Boley}, {Lake}, {Read}, \&
  {Teyssier}}]{2009arXiv0908.1254B}
{Boley} A.~C., {Lake} G., {Read} J., {Teyssier} R., 2009, \apjl, 706, L192

\bibitem[{{Bonnivard} {et~al}\mbox{.}(2015){Bonnivard}, {Combet}, {Daniel},
  {Funk}, {Geringer-Sameth}, {Hinton}, {Maurin}, {Read}, {Sarkar}, {Walker}, \&
  {Wilkinson}}]{2015arXiv150402048B}
{Bonnivard} V. {et~al.}, 2015, \mnras, 453, 849

\bibitem[{{Boylan-Kolchin} {et~al}\mbox{.}(2011){Boylan-Kolchin}, {Bullock}, \&
  {Kaplinghat}}]{2011MNRAS.415L..40B}
{Boylan-Kolchin} M., {Bullock} J.~S., {Kaplinghat} M., 2011, \mnras, 415, L40

\bibitem[{{Breddels} \& {Helmi}(2014)}]{2014ApJ...791L...3B}
{Breddels} M.~A., {Helmi} A., 2014, \apjl, 791, L3

\bibitem[{{Breddels} {et~al}\mbox{.}(2015){Breddels}, {Vera-Ciro}, \&
  {Helmi}}]{2015arXiv150703995B}
{Breddels} M.~A., {Vera-Ciro} C., {Helmi} A., 2015, \apjl, 814, L23

\bibitem[{{Brook} {et~al}\mbox{.}(2014){Brook}, {Di Cintio}, {Knebe},
  {Gottl{\"o}ber}, {Hoffman}, {Yepes}, \&
  {Garrison-Kimmel}}]{2014ApJ...784L..14B}
{Brook} C.~B., {Di Cintio} A., {Knebe} A., {Gottl{\"o}ber} S., {Hoffman} Y.,
  {Yepes} G., {Garrison-Kimmel} S., 2014, \apjl, 784, L14

\bibitem[{{Brooks} {et~al}\mbox{.}(2007){Brooks}, {Governato}, {Booth},
  {Willman}, {Gardner}, {Wadsley}, {Stinson}, \& {Quinn}}]{Brooks07}
{Brooks} A.~M., {Governato} F., {Booth} C.~M., {Willman} B., {Gardner} J.~P.,
  {Wadsley} J., {Stinson} G., {Quinn} T., 2007, \apjl, 655, L17

\bibitem[{{Brooks} {et~al}\mbox{.}(2013){Brooks}, {Kuhlen}, {Zolotov}, \&
  {Hooper}}]{2013ApJ...765...22B}
{Brooks} A.~M., {Kuhlen} M., {Zolotov} A., {Hooper} D., 2013, \apj, 765, 22

\bibitem[{{Bullock} {et~al}\mbox{.}(2001){Bullock}, {Dekel}, {Kolatt},
  {Kravtsov}, {Klypin}, {Porciani}, \& {Primack}}]{2001ApJ...555..240B}
{Bullock} J.~S., {Dekel} A., {Kolatt} T.~S., {Kravtsov} A.~V., {Klypin} A.~A.,
  {Porciani} C., {Primack} J.~R., 2001, \apj, 555, 240

\bibitem[{{Chabrier}(2003)}]{chabrier03}
{Chabrier} G., 2003, \pasp, 115, 763

\bibitem[{{Charbonnier} {et~al}\mbox{.}(2011){Charbonnier}, {Combet}, {Daniel},
  {Funk}, {Hinton}, {Maurin}, {Power}, {Read}, {Sarkar}, {Walker}, \&
  {Wilkinson}}]{2011MNRAS.418.1526C}
{Charbonnier} A. {et~al.}, 2011, \mnras, 418, 1526

\bibitem[{{Cioffi} {et~al}\mbox{.}(1988){Cioffi}, {McKee}, \&
  {Bertschinger}}]{Cioffi1988}
{Cioffi} D.~F., {McKee} C.~F., {Bertschinger} E., 1988, \apj, 334, 252

\bibitem[{{Cole} {et~al}\mbox{.}(2007){Cole}, {Skillman}, {Tolstoy},
  {Gallagher}, {Aparicio}, {Dolphin}, {Gallart}, {Hidalgo}, {Saha}, {Stetson},
  \& {Weisz}}]{2007ApJ...659L..17C}
{Cole} A.~A. {et~al.}, 2007, \apjl, 659, L17

\bibitem[{{Cole} {et~al}\mbox{.}(2014){Cole}, {Weisz}, {Dolphin}, {Skillman},
  {McConnachie}, {Brooks}, \& {Leaman}}]{2014ApJ...795...54C}
{Cole} A.~A., {Weisz} D.~R., {Dolphin} A.~E., {Skillman} E.~D., {McConnachie}
  A.~W., {Brooks} A.~M., {Leaman} R., 2014, \apj, 795, 54

\bibitem[{{Cole} {et~al}\mbox{.}(2012){Cole}, {Dehnen}, {Read}, \&
  {Wilkinson}}]{2012MNRAS.426..601C}
{Cole} D.~R., {Dehnen} W., {Read} J.~I., {Wilkinson} M.~I., 2012, \mnras, 426,
  601

\bibitem[{{Cole} {et~al}\mbox{.}(2011){Cole}, {Dehnen}, \&
  {Wilkinson}}]{2011MNRAS.416.1118C}
{Cole} D.~R., {Dehnen} W., {Wilkinson} M.~I., 2011, \mnras, 416, 1118

\bibitem[{{Collins} {et~al}\mbox{.}(2014){Collins}, {Chapman}, {Rich}, {Ibata},
  {Martin}, {Irwin}, {Bate}, {Lewis}, {Pe{\~n}arrubia}, {Arimoto}, {Casey},
  {Ferguson}, {Koch}, {McConnachie}, \& {Tanvir}}]{2014ApJ...783....7C}
{Collins} M.~L.~M. {et~al.}, 2014, \apj, 783, 7

\bibitem[{{Dalcanton} \& {Hogan}(2001)}]{2001ApJ...561...35D}
{Dalcanton} J.~J., {Hogan} C.~J., 2001, \apj, 561, 35

\bibitem[{{Dekel} \& {Silk}(1986)}]{1986ApJ...303...39D}
{Dekel} A., {Silk} J., 1986, \apj, 303, 39

\bibitem[{{Di Cintio} {et~al}\mbox{.}(2014){Di Cintio}, {Brook}, {Dutton},
  {Macci{\`o}}, {Stinson}, \& {Knebe}}]{2014MNRAS.441.2986D}
{Di Cintio} A., {Brook} C.~B., {Dutton} A.~A., {Macci{\`o}} A.~V., {Stinson}
  G.~S., {Knebe} A., 2014, \mnras, 441, 2986

\bibitem[{{Diemand} {et~al}\mbox{.}(2005){Diemand}, {Madau}, \&
  {Moore}}]{2005MNRAS.364..367D}
{Diemand} J., {Madau} P., {Moore} B., 2005, \mnras, 364, 367

\bibitem[{{Drlica-Wagner} {et~al}\mbox{.}(2015){Drlica-Wagner}, {Albert},
  {Bechtol}, {Wood}, {Strigari}, {S{\'a}nchez-Conde}, {Baldini}, {Essig},
  {Cohen-Tanugi}, {Anderson}, {Bellazzini}, {Bloom}, {Caputo}, {Cecchi},
  {Charles}, {Chiang}, {de Angelis}, {Funk}, {Fusco}, {Gargano}, {Giglietto},
  {Giordano}, {Guiriec}, {Gustafsson}, {Kuss}, {Loparco}, {Lubrano}, {Mirabal},
  {Mizuno}, {Morselli}, {Ohsugi}, {Orlando}, {Persic}, {Rain{\`o}}, {Sehgal},
  {Spada}, {Suson}, {Zaharijas}, {Zimmer}, {Fermi-LAT Collaboration}, {Abbott},
  {Allam}, {Balbinot}, {Bauer}, {Benoit-L{\'e}vy}, {Bernstein}, {Bernstein},
  {Bertin}, {Brooks}, {Buckley-Geer}, {Burke}, {Carnero Rosell}, {Castander},
  {Covarrubias}, {D'Andrea}, {da Costa}, {DePoy}, {Desai}, {Diehl}, {Cunha},
  {Eifler}, {Estrada}, {Evrard}, {Fausti Neto}, {Fernandez}, {Finley},
  {Flaugher}, {Frieman}, {Gaztanaga}, {Gerdes}, {Gruen}, {Gruendl},
  {Gutierrez}, {Honscheid}, {Jain}, {James}, {Jeltema}, {Kent}, {Kron},
  {Kuehn}, {Kuropatkin}, {Lahav}, {Li}, {Luque}, {Maia}, {Makler}, {March},
  {Marshall}, {Martini}, {Merritt}, {Miller}, {Miquel}, {Mohr}, {Neilsen},
  {Nord}, {Ogando}, {Peoples}, {Petravick}, {Pieres}, {Plazas}, {Queiroz},
  {Romer}, {Roodman}, {Rykoff}, {Sako}, {Sanchez}, {Santiago}, {Scarpine},
  {Schubnell}, {Sevilla}, {Smith}, {Soares-Santos}, {Sobreira}, {Suchyta},
  {Swanson}, {Tarle}, {Thaler}, {Thomas}, {Tucker}, {Walker}, {Wechsler},
  {Wester}, {Williams}, {Yanny}, {Zuntz}, \& {DES
  Collaboration}}]{2015arXiv150302632T}
{Drlica-Wagner} A. {et~al.}, 2015, \apjl, 809, L4

\bibitem[{{Dubinski} \& {Carlberg}(1991)}]{1991ApJ...378..496D}
{Dubinski} J., {Carlberg} R.~G., 1991, \apj, 378, 496

\bibitem[{{Dubois} \& {Teyssier}(2008)}]{dubois08}
{Dubois} Y., {Teyssier} R., 2008, \aap, 477, 79

\bibitem[{{Edmunds}(1990)}]{Edmunds1990}
{Edmunds} M.~G., 1990, \mnras, 246, 678

\bibitem[{{Efstathiou}(1992)}]{Efstathiou1992}
{Efstathiou} G., 1992, \mnras, 256, 43P

\bibitem[{{El-Badry} {et~al}\mbox{.}(2015){El-Badry}, {Wetzel}, {Geha},
  {Hopkins}, {Kere{\v s}}, {Chan}, \&
  {Faucher-Gigu{\`e}re}}]{2015arXiv151201235E}
{El-Badry} K., {Wetzel} A.~R., {Geha} M., {Hopkins} P.~F., {Kere{\v s}} D.,
  {Chan} T.~K., {Faucher-Gigu{\`e}re} C.-A., 2015, ArXiv e-prints: 1512.01235

\bibitem[{{El-Zant} {et~al}\mbox{.}(2001){El-Zant}, {Shlosman}, \&
  {Hoffman}}]{2001ApJ...560..636E}
{El-Zant} A., {Shlosman} I., {Hoffman} Y., 2001, \apj, 560, 636

\bibitem[{{Elbert} {et~al}\mbox{.}(2015){Elbert}, {Bullock}, {Garrison-Kimmel},
  {Rocha}, {O{\~n}orbe}, \& {Peter}}]{2014arXiv1412.1477E}
{Elbert} O.~D., {Bullock} J.~S., {Garrison-Kimmel} S., {Rocha} M., {O{\~n}orbe}
  J., {Peter} A.~H.~G., 2015, \mnras, 453, 29

\bibitem[{{Evans} {et~al}\mbox{.}(2014){Evans}, {Heiderman}, \&
  {Vutisalchavakul}}]{evans_etal14}
{Evans}, II N.~J., {Heiderman} A., {Vutisalchavakul} N., 2014, \apj, 782, 114

\bibitem[{{Fisher} {et~al}\mbox{.}(2014){Fisher}, {Bolatto}, {Herrera-Camus},
  {Draine}, {Donaldson}, {Walter}, {Sandstrom}, {Leroy}, {Cannon}, \&
  {Gordon}}]{Fisher2014}
{Fisher} D.~B. {et~al.}, 2014, \nat, 505, 186

\bibitem[{{Flores} \& {Primack}(1994)}]{1994ApJ...427L...1F}
{Flores} R.~A., {Primack} J.~R., 1994, \apjl, 427, L1

\bibitem[{{Garnett}(2002)}]{Garnett02}
{Garnett} D.~R., 2002, \apj, 581, 1019

\bibitem[{{Garrison-Kimmel} {et~al}\mbox{.}(2013){Garrison-Kimmel}, {Rocha},
  {Boylan-Kolchin}, {Bullock}, \& {Lally}}]{2013MNRAS.433.3539G}
{Garrison-Kimmel} S., {Rocha} M., {Boylan-Kolchin} M., {Bullock} J.~S., {Lally}
  J., 2013, \mnras, 433, 3539

\bibitem[{{Gatto} {et~al}\mbox{.}(2013){Gatto}, {Fraternali}, {Read},
  {Marinacci}, {Lux}, \& {Walch}}]{2013MNRAS.433.2749G}
{Gatto} A., {Fraternali} F., {Read} J.~I., {Marinacci} F., {Lux} H., {Walch}
  S., 2013, \mnras, 433, 2749

\bibitem[{{Gatto} {et~al}\mbox{.}(2015){Gatto}, {Walch}, {Low}, {Naab},
  {Girichidis}, {Glover}, {W{\"u}nsch}, {Klessen}, {Clark}, {Baczynski},
  {Peters}, {Ostriker}, {Ib{\'a}{\~n}ez-Mej{\'{\i}}a}, \& {Haid}}]{Gatto2014}
{Gatto} A. {et~al.}, 2015, \mnras, 449, 1057

\bibitem[{{Geringer-Sameth} {et~al}\mbox{.}(2015){Geringer-Sameth}, {Walker},
  {Koushiappas}, {Koposov}, {Belokurov}, {Torrealba}, \&
  {Evans}}]{2015arXiv150302320G}
{Geringer-Sameth} A., {Walker} M.~G., {Koushiappas} S.~M., {Koposov} S.~E.,
  {Belokurov} V., {Torrealba} G., {Evans} N.~W., 2015, Physical Review Letters,
  115, 081101

\bibitem[{{Gnedin}(2000)}]{2000ApJ...542..535G}
{Gnedin} N.~Y., 2000, \apj, 542, 535

\bibitem[{{Gnedin} \& {Draine}(2014)}]{GnedinDraine2014}
{Gnedin} N.~Y., {Draine} B.~T., 2014, \apj, 795, 37

\bibitem[{{Gnedin} \& {Kravtsov}(2011)}]{GnedinKravtsov11}
{Gnedin} N.~Y., {Kravtsov} A.~V., 2011, \apj, 728, 88

\bibitem[{{Gnedin} {et~al}\mbox{.}(2009){Gnedin}, {Tassis}, \&
  {Kravtsov}}]{Gnedin09}
{Gnedin} N.~Y., {Tassis} K., {Kravtsov} A.~V., 2009, \apj, 697, 55

\bibitem[{{Gnedin} \& {Zhao}(2002)}]{2002MNRAS.333..299G}
{Gnedin} O.~Y., {Zhao} H., 2002, \mnras, 333, 299

\bibitem[{{Goerdt} {et~al}\mbox{.}(2010){Goerdt}, {Moore}, {Read}, \&
  {Stadel}}]{2010ApJ...725.1707G}
{Goerdt} T., {Moore} B., {Read} J.~I., {Stadel} J., 2010, \apj, 725, 1707

\bibitem[{{Goerdt} {et~al}\mbox{.}(2006){Goerdt}, {Moore}, {Read}, {Stadel}, \&
  {Zemp}}]{Goerdt:2006rw}
{Goerdt} T., {Moore} B., {Read} J.~I., {Stadel} J., {Zemp} M., 2006, \mnras,
  368, 1073

\bibitem[{{Gonzalez-Samaniego} {et~al}\mbox{.}(2015){Gonzalez-Samaniego},
  {Avila-Reese}, \& {Colin}}]{2015arXiv151203538G}
{Gonzalez-Samaniego} A., {Avila-Reese} V., {Colin} P., 2015, ArXiv e-prints:
  1512.03538

\bibitem[{{Governato} {et~al}\mbox{.}(2010){Governato}, {Brook}, {Mayer},
  {Brooks}, {Rhee}, {Wadsley}, {Jonsson}, {Willman}, {Stinson}, {Quinn}, \&
  {Madau}}]{GovernatoEtAl2010}
{Governato} F. {et~al.}, 2010, \nat, 463, 203

\bibitem[{{Guillet} \& {Teyssier}(2011)}]{GuilletTeyssier2011}
{Guillet} T., {Teyssier} R., 2011, Journal of Computational Physics, 230, 4756

\bibitem[{{Hague} \& {Wilkinson}(2013)}]{2013MNRAS.433.2314H}
{Hague} P.~R., {Wilkinson} M.~I., 2013, \mnras, 433, 2314

\bibitem[{{Hague} \& {Wilkinson}(2014)}]{2014MNRAS.443.3712H}
{Hague} P.~R., {Wilkinson} M.~I., 2014, \mnras, 443, 3712

\bibitem[{{Hidalgo} {et~al}\mbox{.}(2011){Hidalgo}, {Aparicio}, {Skillman},
  {Monelli}, {Gallart}, {Cole}, {Dolphin}, {Weisz}, {Bernard}, {Cassisi},
  {Mayer}, {Stetson}, {Tolstoy}, \& {Ferguson}}]{2011ApJ...730...14H}
{Hidalgo} S.~L. {et~al.}, 2011, \apj, 730, 14

\bibitem[{{Hobbs} {et~al}\mbox{.}(2015){Hobbs}, {Read}, \&
  {Nicola}}]{2015MNRAS.452.3593H}
{Hobbs} A., {Read} J., {Nicola} A., 2015, \mnras, 452, 3593

\bibitem[{{Hockney} \& {Eastwood}(1981)}]{Hockney1981}
{Hockney} R.~W., {Eastwood} J.~W., 1981, {Computer Simulation Using Particles}.
  New York: McGraw-Hill, 1981

\bibitem[{{Hopkins} {et~al}\mbox{.}(2014){Hopkins}, {Kere{\v s}}, {O{\~n}orbe},
  {Faucher-Gigu{\`e}re}, {Quataert}, {Murray}, \& {Bullock}}]{Hopkins2014}
{Hopkins} P.~F., {Kere{\v s}} D., {O{\~n}orbe} J., {Faucher-Gigu{\`e}re} C.-A.,
  {Quataert} E., {Murray} N., {Bullock} J.~S., 2014, \mnras, 445, 581

\bibitem[{{Hopkins} {et~al}\mbox{.}(2013){Hopkins}, {Narayanan}, \&
  {Murray}}]{2013MNRAS.432.2647H}
{Hopkins} P.~F., {Narayanan} D., {Murray} N., 2013, \mnras, 432, 2647

\bibitem[{{Hopkins} {et~al}\mbox{.}(2011){Hopkins}, {Quataert}, \&
  {Murray}}]{2011MNRAS.417..950H}
{Hopkins} P.~F., {Quataert} E., {Murray} N., 2011, \mnras, 417, 950

\bibitem[{{Horiuchi} {et~al}\mbox{.}(2014){Horiuchi}, {Humphrey}, {O{\~n}orbe},
  {Abazajian}, {Kaplinghat}, \& {Garrison-Kimmel}}]{2014PhRvD..89b5017H}
{Horiuchi} S., {Humphrey} P.~J., {O{\~n}orbe} J., {Abazajian} K.~N.,
  {Kaplinghat} M., {Garrison-Kimmel} S., 2014, \prd, 89, 025017

\bibitem[{{Irwin} {et~al}\mbox{.}(2007){Irwin}, {Belokurov}, {Evans},
  {Ryan-Weber}, {de Jong}, {Koposov}, {Zucker}, {Hodgkin}, {Gilmore}, {Prema},
  {Hebb}, {Begum}, {Fellhauer}, {Hewett}, {Kennicutt}, {Wilkinson}, {Bramich},
  {Vidrih}, {Rix}, {Beers}, {Barentine}, {Brewington}, {Harvanek},
  {Krzesinski}, {Long}, {Nitta}, \& {Snedden}}]{2007ApJ...656L..13I}
{Irwin} M.~J. {et~al.}, 2007, \apjl, 656, L13

\bibitem[{{Jardel} \& {Gebhardt}(2013)}]{2013ApJ...775L..30J}
{Jardel} J.~R., {Gebhardt} K., 2013, \apjl, 775, L30

\bibitem[{{Johnson} {et~al}\mbox{.}(2009){Johnson}, {Greif}, {Bromm},
  {Klessen}, \& {Ippolito}}]{2009MNRAS.399...37J}
{Johnson} J.~L., {Greif} T.~H., {Bromm} V., {Klessen} R.~S., {Ippolito} J.,
  2009, \mnras, 399, 37

\bibitem[{{Johnston} {et~al}\mbox{.}(2002){Johnston}, {Spergel}, \&
  {Haydn}}]{2002ApJ...570..656J}
{Johnston} K.~V., {Spergel} D.~N., {Haydn} C., 2002, \apj, 570, 656

\bibitem[{{Karlsson} {et~al}\mbox{.}(2013){Karlsson}, {Bromm}, \&
  {Bland-Hawthorn}}]{2013RvMP...85..809K}
{Karlsson} T., {Bromm} V., {Bland-Hawthorn} J., 2013, Reviews of Modern
  Physics, 85, 809

\bibitem[{{Kauffmann}(2014)}]{2014MNRAS.441.2717K}
{Kauffmann} G., 2014, \mnras, 441, 2717

\bibitem[{{Kaufmann} {et~al}\mbox{.}(2006){Kaufmann}, {Mayer}, {Wadsley},
  {Stadel}, \& {Moore}}]{2006MNRAS.370.1612K}
{Kaufmann} T., {Mayer} L., {Wadsley} J., {Stadel} J., {Moore} B., 2006, \mnras,
  370, 1612

\bibitem[{{Kazantzidis} {et~al}\mbox{.}(2008){Kazantzidis}, {Bullock},
  {Zentner}, {Kravtsov}, \& {Moustakas}}]{2007arXiv0708.1949K}
{Kazantzidis} S., {Bullock} J.~S., {Zentner} A.~R., {Kravtsov} A.~V.,
  {Moustakas} L.~A., 2008, \apj, 688, 254

\bibitem[{{Kazantzidis} {et~al}\mbox{.}(2013){Kazantzidis}, {{\L}okas}, \&
  {Mayer}}]{2013ApJ...764L..29K}
{Kazantzidis} S., {{\L}okas} E.~L., {Mayer} L., 2013, \apjl, 764, L29

\bibitem[{{Kim} \& {Ostriker}(2015)}]{KimOstriker2014}
{Kim} C.-G., {Ostriker} E.~C., 2015, \apj, 802, 99

\bibitem[{{Kimm} {et~al}\mbox{.}(2015){Kimm}, {Cen}, {Devriendt}, {Dubois}, \&
  {Slyz}}]{2015arXiv150105655K}
{Kimm} T., {Cen} R., {Devriendt} J., {Dubois} Y., {Slyz} A., 2015, \mnras, 451,
  2900

\bibitem[{{Kirby} {et~al}\mbox{.}(2014){Kirby}, {Bullock}, {Boylan-Kolchin},
  {Kaplinghat}, \& {Cohen}}]{2014MNRAS.439.1015K}
{Kirby} E.~N., {Bullock} J.~S., {Boylan-Kolchin} M., {Kaplinghat} M., {Cohen}
  J.~G., 2014, \mnras, 439, 1015

\bibitem[{{Kirby} {et~al}\mbox{.}(2013){Kirby}, {Cohen}, {Guhathakurta},
  {Cheng}, {Bullock}, \& {Gallazzi}}]{2013ApJ...779..102K}
{Kirby} E.~N., {Cohen} J.~G., {Guhathakurta} P., {Cheng} L., {Bullock} J.~S.,
  {Gallazzi} A., 2013, \apj, 779, 102

\bibitem[{{Kleyna} {et~al}\mbox{.}(2003){Kleyna}, {Wilkinson}, {Gilmore}, \&
  {Evans}}]{2003ApJ...588L..21K}
{Kleyna} J.~T., {Wilkinson} M.~I., {Gilmore} G., {Evans} N.~W., 2003, \apjl,
  588, L21

\bibitem[{{Klypin} {et~al}\mbox{.}(1999){Klypin}, {Kravtsov}, {Valenzuela}, \&
  {Prada}}]{1999ApJ...522...82K}
{Klypin} A., {Kravtsov} A.~V., {Valenzuela} O., {Prada} F., 1999, \apj, 522, 82

\bibitem[{{Koposov} {et~al}\mbox{.}(2009){Koposov}, {Yoo}, {Rix}, {Weinberg},
  {Macci{\`o}}, \& {Escud{\'e}}}]{2009ApJ...696.2179K}
{Koposov} S.~E., {Yoo} J., {Rix} H.-W., {Weinberg} D.~H., {Macci{\`o}} A.~V.,
  {Escud{\'e}} J.~M., 2009, \apj, 696, 2179

\bibitem[{{Kowalczyk} {et~al}\mbox{.}(2013){Kowalczyk}, {{\L}okas},
  {Kazantzidis}, \& {Mayer}}]{2013MNRAS.431.2796K}
{Kowalczyk} K., {{\L}okas} E.~L., {Kazantzidis} S., {Mayer} L., 2013, \mnras,
  431, 2796

\bibitem[{{Kravtsov}(2003)}]{2003ApJ...590L...1K}
{Kravtsov} A.~V., 2003, \apjl, 590, L1

\bibitem[{{Kravtsov}(2013)}]{2013ApJ...764L..31K}
{Kravtsov} A.~V., 2013, \apjl, 764, L31

\bibitem[{{Kroupa}(2002)}]{2002Sci...295...82K}
{Kroupa} P., 2002, Science, 295, 82

\bibitem[{{Krumholz} {et~al}\mbox{.}(2009){Krumholz}, {McKee}, \&
  {Tumlinson}}]{Krumholz09}
{Krumholz} M.~R., {McKee} C.~F., {Tumlinson} J., 2009, \apj, 699, 850

\bibitem[{{Krumholz} \& {Tan}(2007)}]{krumholztan07}
{Krumholz} M.~R., {Tan} J.~C., 2007, \apj, 654, 304

\bibitem[{{Kuzio de Naray} \& {Kaufmann}(2011)}]{2011MNRAS.414.3617K}
{Kuzio de Naray} R., {Kaufmann} T., 2011, \mnras, 414, 3617

\bibitem[{{Lada} {et~al}\mbox{.}(2010){Lada}, {Lombardi}, \&
  {Alves}}]{lada_etal10}
{Lada} C.~J., {Lombardi} M., {Alves} J.~F., 2010, \apj, 724, 687

\bibitem[{{Lake}(1990)}]{1990Natur.346...39L}
{Lake} G., 1990, \nat, 346, 39

\bibitem[{{Laporte} \& {Pe{\~n}arrubia}(2015)}]{2015MNRAS.449L..90L}
{Laporte} C.~F.~P., {Pe{\~n}arrubia} J., 2015, \mnras, 449, L90

\bibitem[{{Leaman} {et~al}\mbox{.}(2012){Leaman}, {Venn}, {Brooks},
  {Battaglia}, {Cole}, {Ibata}, {Irwin}, {McConnachie}, {Mendel}, \&
  {Tolstoy}}]{2012ApJ...750...33L}
{Leaman} R. {et~al.}, 2012, \apj, 750, 33

\bibitem[{{Lee} {et~al}\mbox{.}(1999){Lee}, {Aparicio}, {Tikonov}, {Byun}, \&
  {Kim}}]{1999AJ....118..853L}
{Lee} M.~G., {Aparicio} A., {Tikonov} N., {Byun} Y.-I., {Kim} E., 1999, \aj,
  118, 853

\bibitem[{{Leitherer} {et~al}\mbox{.}(1999){Leitherer}, {Schaerer}, {Goldader},
  {Gonz{\'a}lez Delgado}, {Robert}, {Kune}, {de Mello}, {Devost}, \&
  {Heckman}}]{Leitherer1999}
{Leitherer} C. {et~al.}, 1999, \apjs, 123, 3

\bibitem[{{{\L}okas} {et~al}\mbox{.}(2012){{\L}okas}, {Kazantzidis}, \&
  {Mayer}}]{2012ApJ...751L..15L}
{{\L}okas} E.~L., {Kazantzidis} S., {Mayer} L., 2012, \apjl, 751, L15

\bibitem[{{Lux} {et~al}\mbox{.}(2010){Lux}, {Read}, \&
  {Lake}}]{2010arXiv1001.1731L}
{Lux} H., {Read} J.~I., {Lake} G., 2010, \mnras, 406, 2312

\bibitem[{{Macci{\`o}} {et~al}\mbox{.}(2007){Macci{\`o}}, {Dutton}, {van den
  Bosch}, {Moore}, {Potter}, \& {Stadel}}]{2007MNRAS.378...55M}
{Macci{\`o}} A.~V., {Dutton} A.~A., {van den Bosch} F.~C., {Moore} B., {Potter}
  D., {Stadel} J., 2007, \mnras, 378, 55

\bibitem[{{Macci{\`o}} {et~al}\mbox{.}(2012){Macci{\`o}}, {Paduroiu},
  {Anderhalden}, {Schneider}, \& {Moore}}]{2012MNRAS.424.1105M}
{Macci{\`o}} A.~V., {Paduroiu} S., {Anderhalden} D., {Schneider} A., {Moore}
  B., 2012, \mnras, 424, 1105

\bibitem[{{Madau} {et~al}\mbox{.}(2014){Madau}, {Shen}, \&
  {Governato}}]{2014ApJ...789L..17M}
{Madau} P., {Shen} S., {Governato} F., 2014, \apjl, 789, L17

\bibitem[{{Malyshev} {et~al}\mbox{.}(2014){Malyshev}, {Neronov}, \&
  {Eckert}}]{2014PhRvD..90j3506M}
{Malyshev} D., {Neronov} A., {Eckert} D., 2014, \prd, 90, 103506

\bibitem[{{Martin} {et~al}\mbox{.}(2008){Martin}, {de Jong}, \&
  {Rix}}]{2008ApJ...684.1075M}
{Martin} N.~F., {de Jong} J.~T.~A., {Rix} H.-W., 2008, \apj, 684, 1075

\bibitem[{{Martizzi} {et~al}\mbox{.}(2015){Martizzi}, {Faucher-Gigu{\`e}re}, \&
  {Quataert}}]{Martizzi2014}
{Martizzi} D., {Faucher-Gigu{\`e}re} C.-A., {Quataert} E., 2015, \mnras, 450,
  504

\bibitem[{{Mashchenko} {et~al}\mbox{.}(2008){Mashchenko}, {Wadsley}, \&
  {Couchman}}]{2008Sci...319..174M}
{Mashchenko} S., {Wadsley} J., {Couchman} H.~M.~P., 2008, Science, 319, 174

\bibitem[{{Maxwell} {et~al}\mbox{.}(2015){Maxwell}, {Wadsley}, \&
  {Couchman}}]{2015arXiv150500825M}
{Maxwell} A.~J., {Wadsley} J., {Couchman} H.~M.~P., 2015, \apj, 806, 229

\bibitem[{{Mayer} {et~al}\mbox{.}(2001){Mayer}, {Governato}, {Colpi}, {Moore},
  {Quinn}, {Wadsley}, {Stadel}, \& {Lake}}]{2001ApJ...547L.123M}
{Mayer} L., {Governato} F., {Colpi} M., {Moore} B., {Quinn} T., {Wadsley} J.,
  {Stadel} J., {Lake} G., 2001, \apjl, 547, L123

\bibitem[{{McConnachie}(2012)}]{2012AJ....144....4M}
{McConnachie} A.~W., 2012, \aj, 144, 4

\bibitem[{{McConnachie} {et~al}\mbox{.}(2006){McConnachie}, {Arimoto}, {Irwin},
  \& {Tolstoy}}]{2006MNRAS.373..715M}
{McConnachie} A.~W., {Arimoto} N., {Irwin} M., {Tolstoy} E., 2006, \mnras, 373,
  715

\bibitem[{{McQuinn} {et~al}\mbox{.}(2015){McQuinn}, {Lelli}, {Skillman},
  {Dolphin}, {McGaugh}, \& {Williams}}]{2015MNRAS.450.3886M}
{McQuinn} K.~B.~W., {Lelli} F., {Skillman} E.~D., {Dolphin} A.~E., {McGaugh}
  S.~S., {Williams} B.~F., 2015, \mnras, 450, 3886

\bibitem[{{Metcalf} \& {Zhao}(2002)}]{2002ApJ...567L...5M}
{Metcalf} R.~B., {Zhao} H., 2002, \apjl, 567, L5

\bibitem[{{Moore}(1994)}]{1994Natur.370..629M}
{Moore} B., 1994, \nat, 370, 629

\bibitem[{{Moore} {et~al}\mbox{.}(2006){Moore}, {Diemand}, {Madau}, {Zemp}, \&
  {Stadel}}]{2005astro.ph.10370M}
{Moore} B., {Diemand} J., {Madau} P., {Zemp} M., {Stadel} J., 2006, \mnras,
  368, 563

\bibitem[{{Moore} {et~al}\mbox{.}(1999){Moore}, {Ghigna}, {Governato}, {Lake},
  {Quinn}, {Stadel}, \& {Tozzi}}]{1999ApJ...524L..19M}
{Moore} B., {Ghigna} S., {Governato} F., {Lake} G., {Quinn} T., {Stadel} J.,
  {Tozzi} P., 1999, \apjl, 524, L19

\bibitem[{{Moster} {et~al}\mbox{.}(2010){Moster}, {Somerville}, {Maulbetsch},
  {van den Bosch}, {Macci{\`o}}, {Naab}, \& {Oser}}]{2010ApJ...710..903M}
{Moster} B.~P., {Somerville} R.~S., {Maulbetsch} C., {van den Bosch} F.~C.,
  {Macci{\`o}} A.~V., {Naab} T., {Oser} L., 2010, \apj, 710, 903

\bibitem[{{Murray}(2011)}]{Murray2011b}
{Murray} N., 2011, \apj, 729, 133

\bibitem[{{Nakamura} \& {Umemura}(2001)}]{2001ApJ...548...19N}
{Nakamura} F., {Umemura} M., 2001, \apj, 548, 19

\bibitem[{{Navarro} {et~al}\mbox{.}(1996{\natexlab{a}}){Navarro}, {Eke}, \&
  {Frenk}}]{1996MNRAS.283L..72N}
{Navarro} J.~F., {Eke} V.~R., {Frenk} C.~S., 1996{\natexlab{a}}, \mnras, 283,
  L72

\bibitem[{{Navarro} {et~al}\mbox{.}(1996{\natexlab{b}}){Navarro}, {Frenk}, \&
  {White}}]{1996ApJ...462..563N}
{Navarro} J.~F., {Frenk} C.~S., {White} S.~D.~M., 1996{\natexlab{b}}, \apj,
  462, 563

\bibitem[{{Nipoti} \& {Binney}(2015)}]{2015MNRAS.446.1820N}
{Nipoti} C., {Binney} J., 2015, \mnras, 446, 1820

\bibitem[{{O{\~n}orbe} {et~al}\mbox{.}(2015){O{\~n}orbe}, {Boylan-Kolchin},
  {Bullock}, {Hopkins}, {Kere{\v s}}, {Faucher-Gigu{\`e}re}, {Quataert}, \&
  {Murray}}]{2015arXiv150202036O}
{O{\~n}orbe} J., {Boylan-Kolchin} M., {Bullock} J.~S., {Hopkins} P.~F.,
  {Kere{\v s}} D., {Faucher-Gigu{\`e}re} C.-A., {Quataert} E., {Murray} N.,
  2015, \mnras, 454, 2092

\bibitem[{{Ogiya} \& {Burkert}(2015)}]{2015MNRAS.446.2363O}
{Ogiya} G., {Burkert} A., 2015, \mnras, 446, 2363

\bibitem[{{Oh} {et~al}\mbox{.}(2011){Oh}, {Brook}, {Governato}, {Brinks},
  {Mayer}, {de Blok}, {Brooks}, \& {Walter}}]{2011AJ....142...24O}
{Oh} S.-H., {Brook} C., {Governato} F., {Brinks} E., {Mayer} L., {de Blok}
  W.~J.~G., {Brooks} A., {Walter} F., 2011, \aj, 142, 24

\bibitem[{{Oh} {et~al}\mbox{.}(2015){Oh}, {Hunter}, {Brinks}, {Elmegreen},
  {Schruba}, {Walter}, {Rupen}, {Young}, {Simpson}, {Johnson}, {Herrmann},
  {Ficut-Vicas}, {Cigan}, {Heesen}, {Ashley}, \& {Zhang}}]{2015AJ....149..180O}
{Oh} S.-H. {et~al.}, 2015, \aj, 149, 180

\bibitem[{{Pagel} \& {Patchett}(1975)}]{PagelPatchett1975}
{Pagel} B.~E.~J., {Patchett} B.~E., 1975, \mnras, 172, 13

\bibitem[{{Pe{\~n}arrubia} {et~al}\mbox{.}(2010){Pe{\~n}arrubia}, {Benson},
  {Walker}, {Gilmore}, {McConnachie}, \& {Mayer}}]{2010MNRAS.406.1290P}
{Pe{\~n}arrubia} J., {Benson} A.~J., {Walker} M.~G., {Gilmore} G.,
  {McConnachie} A.~W., {Mayer} L., 2010, \mnras, 406, 1290

\bibitem[{{Pe{\~n}arrubia} {et~al}\mbox{.}(2012){Pe{\~n}arrubia}, {Pontzen},
  {Walker}, \& {Koposov}}]{2012ApJ...759L..42P}
{Pe{\~n}arrubia} J., {Pontzen} A., {Walker} M.~G., {Koposov} S.~E., 2012,
  \apjl, 759, L42

\bibitem[{{Planck Collaboration} {et~al}\mbox{.}(2014){Planck Collaboration},
  {Ade}, {Aghanim}, {Armitage-Caplan}, {Arnaud}, {Ashdown}, {Atrio-Barandela},
  {Aumont}, {Baccigalupi}, {Banday}, \& et~al.}]{2013arXiv1303.5076P}
{Planck Collaboration} {et~al.}, 2014, \aap, 571, A16

\bibitem[{{Pontzen} \& {Governato}(2012)}]{2012MNRAS.421.3464P}
{Pontzen} A., {Governato} F., 2012, \mnras, 421, 3464

\bibitem[{{Pontzen} \& {Governato}(2014)}]{2014Natur.506..171P}
{Pontzen} A., {Governato} F., 2014, \nat, 506, 171

\bibitem[{{Pontzen} {et~al}\mbox{.}(2015){Pontzen}, {Read}, {Teyssier},
  {Governato}, {Gualandris}, {Roth}, \& {Devriendt}}]{2015arXiv150207356P}
{Pontzen} A., {Read} J.~I., {Teyssier} R., {Governato} F., {Gualandris} A.,
  {Roth} N., {Devriendt} J., 2015, \mnras, 451, 1366

\bibitem[{{Pontzen} {et~al}\mbox{.}(2013){Pontzen}, {Ro{\v s}kar}, {Stinson},
  \& {Woods}}]{2013ascl.soft05002P}
{Pontzen} A., {Ro{\v s}kar} R., {Stinson} G., {Woods} R., 2013, {pynbody:
  N-Body/SPH analysis for python}. Astrophysics Source Code Library

\bibitem[{Raiteri {et~al}\mbox{.}(1996)Raiteri, Villata, \&
  Navarro}]{Raiteri1996}
Raiteri C.~M., Villata M., Navarro J.~F., 1996, A{\&}A, 315, 105

\bibitem[{{Read} \& {Gilmore}(2005)}]{2005MNRAS.356..107R}
{Read} J.~I., {Gilmore} G., 2005, \mnras, 356, 107

\bibitem[{{Read} {et~al}\mbox{.}(2006{\natexlab{a}}){Read}, {Goerdt}, {Moore},
  {Pontzen}, {Stadel}, \& {Lake}}]{2006astro.ph..6636R}
{Read} J.~I., {Goerdt} T., {Moore} B., {Pontzen} A.~P., {Stadel} J., {Lake} G.,
  2006{\natexlab{a}}, \mnras, 373, 1451

\bibitem[{{Read} {et~al}\mbox{.}(2008){Read}, {Lake}, {Agertz}, \&
  {Debattista}}]{2008MNRAS.389.1041R}
{Read} J.~I., {Lake} G., {Agertz} O., {Debattista} V.~P., 2008, \mnras, 389,
  1041

\bibitem[{{Read} {et~al}\mbox{.}(2006{\natexlab{b}}){Read}, {Pontzen}, \&
  {Viel}}]{2006MNRAS.371..885R}
{Read} J.~I., {Pontzen} A.~P., {Viel} M., 2006{\natexlab{b}}, \mnras, 371, 885

\bibitem[{{Read} {et~al}\mbox{.}(2006{\natexlab{c}}){Read}, {Wilkinson},
  {Evans}, {Gilmore}, \& {Kleyna}}]{2006MNRAS.tmp..153R}
{Read} J.~I., {Wilkinson} M.~I., {Evans} N.~W., {Gilmore} G., {Kleyna} J.~T.,
  2006{\natexlab{c}}, \mnras, 367, 387

\bibitem[{{Read} {et~al}\mbox{.}(2006{\natexlab{d}}){Read}, {Wilkinson},
  {Evans}, {Gilmore}, \& {Kleyna}}]{2006MNRAS.366..429R}
{Read} J.~I., {Wilkinson} M.~I., {Evans} N.~W., {Gilmore} G., {Kleyna} J.~T.,
  2006{\natexlab{d}}, \mnras, 366, 429

\bibitem[{{Richardson} \& {Fairbairn}(2014)}]{2014MNRAS.441.1584R}
{Richardson} T., {Fairbairn} M., 2014, \mnras, 441, 1584

\bibitem[{{Ricotti}(2009)}]{2009MNRAS.392L..45R}
{Ricotti} M., 2009, \mnras, 392, L45

\bibitem[{{Ricotti} {et~al}\mbox{.}(2008){Ricotti}, {Gnedin}, \&
  {Shull}}]{Ricotti2008}
{Ricotti} M., {Gnedin} N.~Y., {Shull} J.~M., 2008, \apj, 685, 21

\bibitem[{{Rocha} {et~al}\mbox{.}(2013){Rocha}, {Peter}, {Bullock},
  {Kaplinghat}, {Garrison-Kimmel}, {O{\~n}orbe}, \&
  {Moustakas}}]{2013MNRAS.430...81R}
{Rocha} M., {Peter} A.~H.~G., {Bullock} J.~S., {Kaplinghat} M.,
  {Garrison-Kimmel} S., {O{\~n}orbe} J., {Moustakas} L.~A., 2013, \mnras, 430,
  81

\bibitem[{{Romeo} {et~al}\mbox{.}(2008){Romeo}, {Agertz}, {Moore}, \&
  {Stadel}}]{Romeo08}
{Romeo} A.~B., {Agertz} O., {Moore} B., {Stadel} J., 2008, \apj, 686, 1

\bibitem[{{Rosen} \& {Bregman}(1995)}]{rosenbregman95}
{Rosen} A., {Bregman} J.~N., 1995, \apj, 440, 634

\bibitem[{{Ryan-Weber} {et~al}\mbox{.}(2008){Ryan-Weber}, {Begum}, {Oosterloo},
  {Pal}, {Irwin}, {Belokurov}, {Evans}, \& {Zucker}}]{2008MNRAS.384..535R}
{Ryan-Weber} E.~V., {Begum} A., {Oosterloo} T., {Pal} S., {Irwin} M.~J.,
  {Belokurov} V., {Evans} N.~W., {Zucker} D.~B., 2008, \mnras, 384, 535

\bibitem[{{Saitoh} {et~al}\mbox{.}(2008){Saitoh}, {Daisaka}, {Kokubo},
  {Makino}, {Okamoto}, {Tomisaka}, {Wada}, \& {Yoshida}}]{2008PASJ...60..667S}
{Saitoh} T.~R., {Daisaka} H., {Kokubo} E., {Makino} J., {Okamoto} T.,
  {Tomisaka} K., {Wada} K., {Yoshida} N., 2008, \pasj, 60, 667

\bibitem[{{Shao} {et~al}\mbox{.}(2013){Shao}, {Gao}, {Theuns}, \&
  {Frenk}}]{2013MNRAS.430.2346S}
{Shao} S., {Gao} L., {Theuns} T., {Frenk} C.~S., 2013, \mnras, 430, 2346

\bibitem[{{Shibuya} {et~al}\mbox{.}(2015){Shibuya}, {Ouchi}, \&
  {Harikane}}]{2015ApJS..219...15S}
{Shibuya} T., {Ouchi} M., {Harikane} Y., 2015, \apjs, 219, 15

\bibitem[{{Simon} {et~al}\mbox{.}(2005){Simon}, {Bolatto}, {Leroy}, {Blitz}, \&
  {Gates}}]{2005ApJ...621..757S}
{Simon} J.~D., {Bolatto} A.~D., {Leroy} A., {Blitz} L., {Gates} E.~L., 2005,
  \apj, 621, 757

\bibitem[{{Simon} \& {Geha}(2007)}]{2007arXiv0706.0516S}
{Simon} J.~D., {Geha} M., 2007, \apj, 670, 313

\bibitem[{{Simpson} {et~al}\mbox{.}(2015){Simpson}, {Bryan}, {Hummels}, \&
  {Ostriker}}]{Simpson2014}
{Simpson} C.~M., {Bryan} G.~L., {Hummels} C., {Ostriker} J.~P., 2015, \apj,
  809, 69

\bibitem[{{Skillman} {et~al}\mbox{.}(2014){Skillman}, {Hidalgo}, {Weisz},
  {Monelli}, {Gallart}, {Aparicio}, {Bernard}, {Boylan-Kolchin}, {Cassisi},
  {Cole}, {Dolphin}, {Ferguson}, {Mayer}, {Navarro}, {Stetson}, \&
  {Tolstoy}}]{2014ApJ...786...44S}
{Skillman} E.~D. {et~al.}, 2014, \apj, 786, 44

\bibitem[{{Springel} {et~al}\mbox{.}(2008){Springel}, {Wang}, {Vogelsberger},
  {Ludlow}, {Jenkins}, {Helmi}, {Navarro}, {Frenk}, \&
  {White}}]{2008MNRAS.391.1685S}
{Springel} V. {et~al.}, 2008, \mnras, 391, 1685

\bibitem[{{Stadel} {et~al}\mbox{.}(2009){Stadel}, {Potter}, {Moore}, {Diemand},
  {Madau}, {Zemp}, {Kuhlen}, \& {Quilis}}]{2009MNRAS.398L..21S}
{Stadel} J., {Potter} D., {Moore} B., {Diemand} J., {Madau} P., {Zemp} M.,
  {Kuhlen} M., {Quilis} V., 2009, \mnras, 398, L21

\bibitem[{{Strigari} {et~al}\mbox{.}(2014){Strigari}, {Frenk}, \&
  {White}}]{2014arXiv1406.6079S}
{Strigari} L.~E., {Frenk} C.~S., {White} S.~D.~M., 2014, ArXiv e-prints:
  1406.6079

\bibitem[{{Strigari} {et~al}\mbox{.}(2007){Strigari}, {Kaplinghat}, \&
  {Bullock}}]{2007PhRvD..75f1303S}
{Strigari} L.~E., {Kaplinghat} M., {Bullock} J.~S., 2007, \prd, 75, 061303

\bibitem[{{Susa} \& {Umemura}(2004)}]{2004ApJ...600....1S}
{Susa} H., {Umemura} M., 2004, \apj, 600, 1

\bibitem[{{Sutherland} \& {Dopita}(1993)}]{sutherlanddopita93}
{Sutherland} R.~S., {Dopita} M.~A., 1993, \apjs, 88, 253

\bibitem[{{Tassis} {et~al}\mbox{.}(2008){Tassis}, {Kravtsov}, \&
  {Gnedin}}]{Tassis08}
{Tassis} K., {Kravtsov} A.~V., {Gnedin} N.~Y., 2008, \apj, 672, 888

\bibitem[{{Teyssier}(2002)}]{teyssier02}
{Teyssier} R., 2002, \aap, 385, 337

\bibitem[{{Teyssier} {et~al}\mbox{.}(2013){Teyssier}, {Pontzen}, {Dubois}, \&
  {Read}}]{2013MNRAS.429.3068T}
{Teyssier} R., {Pontzen} A., {Dubois} Y., {Read} J.~I., 2013, \mnras, 429, 3068

\bibitem[{{Thornton} {et~al}\mbox{.}(1998){Thornton}, {Gaudlitz}, {Janka}, \&
  {Steinmetz}}]{Thornton1998}
{Thornton} K., {Gaudlitz} M., {Janka} H.-T., {Steinmetz} M., 1998, \apj, 500,
  95

\bibitem[{{Tollet} {et~al}\mbox{.}(2016){Tollet}, {Macci{\`o}}, {Dutton},
  {Stinson}, {Wang}, {Penzo}, {Gutcke}, {Buck}, {Kang}, {Brook}, {Di Cintio},
  {Keller}, \& {Wadsley}}]{2015arXiv150703590T}
{Tollet} E. {et~al.}, 2016, \mnras, 456, 3542

\bibitem[{{Tremonti} {et~al}\mbox{.}(2004){Tremonti}, {Heckman}, {Kauffmann},
  {Brinchmann}, {Charlot}, {White}, {Seibert}, {Peng}, {Schlegel}, {Uomoto},
  {Fukugita}, \& {Brinkmann}}]{Tremonti2004}
{Tremonti} C.~A. {et~al.}, 2004, \apj, 613, 898

\bibitem[{{Trujillo-Gomez} {et~al}\mbox{.}(2015){Trujillo-Gomez}, {Klypin},
  {Col{\'{\i}}n}, {Ceverino}, {Arraki}, \& {Primack}}]{2015MNRAS.446.1140T}
{Trujillo-Gomez} S., {Klypin} A., {Col{\'{\i}}n} P., {Ceverino} D., {Arraki}
  K.~S., {Primack} J., 2015, \mnras, 446, 1140

\bibitem[{{Ural} {et~al}\mbox{.}(2015){Ural}, {Wilkinson}, {Read}, \&
  {Walker}}]{2015NatCo...6E7599U}
{Ural} U., {Wilkinson} M.~I., {Read} J.~I., {Walker} M.~G., 2015, Nature
  Communications, 6, 7599

\bibitem[{{van den Bergh}(1959)}]{1959PDDO....2..147V}
{van den Bergh} S., 1959, Publications of the David Dunlap Observatory, 2, 147

\bibitem[{{Verbeke} {et~al}\mbox{.}(2015){Verbeke}, {Vandenbroucke}, \& {De
  Rijcke}}]{2015arXiv151101484V}
{Verbeke} R., {Vandenbroucke} B., {De Rijcke} S., 2015, \apj, 815, 85

\bibitem[{{Villaescusa-Navarro} \& {Dalal}(2011)}]{2011JCAP...03..024V}
{Villaescusa-Navarro} F., {Dalal} N., 2011, JCAP, 3, 24

\bibitem[{{Walker} {et~al}\mbox{.}(2009){Walker}, {Mateo}, {Olszewski},
  {Pe{\~n}arrubia}, {Wyn Evans}, \& {Gilmore}}]{2009ApJ...704.1274W}
{Walker} M.~G., {Mateo} M., {Olszewski} E.~W., {Pe{\~n}arrubia} J., {Wyn Evans}
  N., {Gilmore} G., 2009, \apj, 704, 1274

\bibitem[{{Walker} \& {Pe{\~n}arrubia}(2011)}]{2011ApJ...742...20W}
{Walker} M.~G., {Pe{\~n}arrubia} J., 2011, \apj, 742, 20

\bibitem[{{Weisz} {et~al}\mbox{.}(2014){Weisz}, {Dolphin}, {Skillman},
  {Holtzman}, {Gilbert}, {Dalcanton}, \& {Williams}}]{2014ApJ...789..148W}
{Weisz} D.~R., {Dolphin} A.~E., {Skillman} E.~D., {Holtzman} J., {Gilbert}
  K.~M., {Dalcanton} J.~J., {Williams} B.~F., 2014, \apj, 789, 148

\bibitem[{{Weisz} {et~al}\mbox{.}(2012{\natexlab{a}}){Weisz}, {Johnson},
  {Johnson}, {Skillman}, {Lee}, {Kennicutt}, {Calzetti}, {van Zee}, {Bothwell},
  {Dalcanton}, {Dale}, \& {Williams}}]{2012ApJ...744...44W}
{Weisz} D.~R. {et~al.}, 2012{\natexlab{a}}, \apj, 744, 44

\bibitem[{{Weisz} {et~al}\mbox{.}(2012{\natexlab{b}}){Weisz}, {Zucker},
  {Dolphin}, {Martin}, {de Jong}, {Holtzman}, {Dalcanton}, {Gilbert},
  {Williams}, {Bell}, {Belokurov}, \& {Wyn Evans}}]{2012ApJ...748...88W}
{Weisz} D.~R. {et~al.}, 2012{\natexlab{b}}, \apj, 748, 88

\bibitem[{{Wetzel} \& {Nagai}(2015)}]{2015ApJ...808...40W}
{Wetzel} A.~R., {Nagai} D., 2015, \apj, 808, 40

\bibitem[{{White} \& {Rees}(1978)}]{1978MNRAS.183..341W}
{White} S.~D.~M., {Rees} M.~J., 1978, \mnras, 183, 341

\bibitem[{{Wise} \& {Abel}(2007)}]{2007ApJ...665..899W}
{Wise} J.~H., {Abel} T., 2007, \apj, 665, 899

\bibitem[{{Wise} {et~al}\mbox{.}(2012){Wise}, {Turk}, {Norman}, \&
  {Abel}}]{2012ApJ...745...50W}
{Wise} J.~H., {Turk} M.~J., {Norman} M.~L., {Abel} T., 2012, \apj, 745, 50

\bibitem[{{Wolf} {et~al}\mbox{.}(2010){Wolf}, {Martinez}, {Bullock},
  {Kaplinghat}, {Geha}, {Mu{\~n}oz}, {Simon}, \& {Avedo}}]{2010MNRAS.406.1220W}
{Wolf} J., {Martinez} G.~D., {Bullock} J.~S., {Kaplinghat} M., {Geha} M.,
  {Mu{\~n}oz} R.~R., {Simon} J.~D., {Avedo} F.~F., 2010, \mnras, 406, 1220

\bibitem[{{Yoshida} {et~al}\mbox{.}(2003){Yoshida}, {Abel}, {Hernquist}, \&
  {Sugiyama}}]{2003ApJ...592..645Y}
{Yoshida} N., {Abel} T., {Hernquist} L., {Sugiyama} N., 2003, \apj, 592, 645

\bibitem[{{Zhang} {et~al}\mbox{.}(2012){Zhang}, {Hunter}, {Elmegreen}, {Gao},
  \& {Schruba}}]{2012AJ....143...47Z}
{Zhang} H.-X., {Hunter} D.~A., {Elmegreen} B.~G., {Gao} Y., {Schruba} A., 2012,
  \aj, 143, 47

\bibitem[{{Zolotov} {et~al}\mbox{.}(2012){Zolotov}, {Brooks}, {Willman},
  {Governato}, {Pontzen}, {Christensen}, {Dekel}, {Quinn}, {Shen}, \&
  {Wadsley}}]{2012ApJ...761...71Z}
{Zolotov} A. {et~al.}, 2012, \apj, 761, 71

\end{thebibliography}

\end{document}